\documentclass[12pt, preprint]{aastex}
\usepackage{bm, graphicx, subfigure, amsmath, morefloats}
\usepackage{color}
\newcommand{\sectionname}{Section}

\newcommand{\figurenames}{\figurename s}
\newcommand{\tc}{\textsl{The~Cannon}} 
\newcommand{\apogee}{\textsl{APOGEE}} 
 
\newcommand{\galah}{\textsl{GALAH}}
\newcommand{\segue}{\textsl{SEGUE}}  
\newcommand{\aspcap}{\textsl{ASPCAP}} 
\newcommand{\gaiaeso}{\textsl{Gaia-ESO}} 
 
\newcommand{\rave}{\textsl{RAVE}} 
\newcommand{\matisse}{\textsl{MATISSE}} 
\newcommand{\rotwarn}{\texttt{ROTATION WARNING}} 
\newcommand{\badstar}{\texttt{STAR BAD}} 
\newcommand{\aspcapflag}{\texttt{ASPCAPFLAG}} 
 
\newcommand{\lamost}{\textsl{LAMOST}} 
\newcommand{\aspcapstar}{\textsl{aspcapStar}} 
\newcommand{\apstar}{\textsl{apStar}} 

\newcommand{\set}[1]{\bm{#1}}
\newcommand{\starlabel}{\ell}
\newcommand{\starlabelvec}{\set{\starlabel}}
\newcommand{\mean}[1]{\overline{#1}}
\newcommand{\given}{\,|\,}
\newcommand{\teff}{\mbox{$\rm T_{eff}$}}
\newcommand{\kms}{\mbox{$\rm kms^{-1}$}}
\newcommand{\feh}{\mbox{$\rm [Fe/H]$}}
\newcommand{\xfe}{\mbox{$\rm [X/Fe]$}}
\newcommand{\alphafe}{\mbox{$\rm [\alpha/Fe]$}}
\newcommand{\mh}{\mbox{$\rm [M/H]$}}
\newcommand{\logg}{\mbox{$\rm \log g$}}
\newcommand{\noise}{\sigma_{n\lambda}}
\newcommand{\scatter}{s_{\lambda}}
\newcommand{\pix}{\mathrm{pix}}
\newcommand{\rfn}{\mathrm{ref}}

\begin{document}

\title{\tc:\\ A data-driven approach to stellar label determination}
\author{M.~Ness\altaffilmark{1},  
David~W.~Hogg\altaffilmark{1,2,3}, 
H.-W.~Rix\altaffilmark{1}, 
Anna.~Y.~Q.~Ho\altaffilmark{1}, 
G.~Zasowski\altaffilmark{4,5}}
\altaffiltext{1}{Max-Planck-Institut f\"ur Astronomie, K\"onigstuhl 17, D-69117 Heidelberg, Germany}
\altaffiltext{2}{Center for Cosmology and Particle Physics, Department of Phyics,
             New York University, 4 Washington Pl., room 424, New York, NY, 10003, USA}
\altaffiltext{3}{Center for Data Science, New York University, 726 Broadway, 7th Floor, New York, NY 10003, USA}
\altaffiltext{4}{NSF Astronomy and Astrophysics Postdoctoral Fellow}
\altaffiltext{5}{Department of Physics \& Astronomy, Johns Hopkins University, Baltimore, MD, 21218, USA}
\email{ness@mpia.de}

\begin{abstract}%
New spectroscopic surveys offer the promise of stellar
parameters and abundances (`stellar labels') for hundreds of thousands
of stars; this poses a formidable spectral modeling challenge. In many cases, there is a sub-set of \emph{reference
objects} for which the stellar labels are known with high(er)
fidelity. We take advantage of this with \tc , a new data-driven
approach for determining stellar labels from spectroscopic data.
\tc\ learns from the `known' labels of reference stars 
how the continuum-normalized spectra depend on these labels
by fitting a flexible model at each wavelength;
then,  \tc\ uses this model to derive labels for the remaining survey stars.
We illustrate \tc\ by training the model 
 on only 542 stars in 19 clusters as reference objects, with \teff, \logg\ and \feh\ as the labels,
and then applying it to the spectra of 55,000 stars from
\apogee\ DR10.
\tc\ is very accurate. Its stellar labels compare well to the
stars for which \apogee\ pipeline (\aspcap) labels are provided in DR10, with 
\textit{rms} differences that are basically identical to the stated \aspcap\ uncertainties.
Beyond the reference labels, \tc\ makes no use of
stellar models nor any line-list, but needs a set of reference objects that span label-space.
\tc\  performs well at lower signal-to-noise, as it delivers comparably 
good labels even at one ninth the \apogee\ observing time.
We discuss the limitations of \tc\ and its future potential, particularly, 
to bring different spectroscopic surveys onto a consistent
scale of stellar labels.
\end{abstract}


\section{Introduction}\label{sec:Intro}

The vast spectroscopic stellar surveys of recent years (e.g., \segue\ \citep{Beers2006}, \rave\ \citep{Steinmetz2006}, \lamost\ \citep{Newberg2012}, \apogee\ \citep{Majewski2012}, \gaiaeso\ \citep{Gilmore2012}, \galah\ \citep{Freeman2012}) hold tremendous astrophysical promise, but at the same time present formidable data analysis and modeling challenges. 
One of these challenges lies in consistently and accurately determining what we call ``stellar labels'', that is, stellar parameters and element abundances, from survey spectra. 
These labels are usually determined from comparison of the data with synthetic model spectra, with approaches often 
customized specifically to the particular wavelength region of a given survey \citep[e.g.,][]{ Lee2006, Boeche2011, Liu2014, Meszaros2013, SM2014, BJ2013}. 

The stellar photosphere models that are relied upon for stellar label determination have physical ingredients that are incomplete and simplified. 
For computational feasibility, almost always 1D stellar photosphere models are adopted for large surveys, often assumed to be in local thermal equilibrium; these approximations are both severe. 
In many cases, the model spectra do not account for all relevant molecular opacities, for convection, stellar winds, and the chromosphere. 
As a consequence, it happens that different research groups obtain discrepant results for same stars, resulting from analysis across different wavelength regions and different input assumptions and methods used \citep[e.g.,][]{Hinkel2014, Jofre2014, AP1999}. Even when the input assumptions are held fixed, differences in the employed analysis methods lead to substantial differences in assigned labels \citep[e.g.][]{SM2014}.

Stellar labels are commonly determined by fitting a grid of model spectra (with known labels) to the data using some minimisation technique, often restricted to a masked portion of the spectrum that is focused on the absorption line (regions) deemed to be most reliable or relevant. 
Stated minimal signal-to-noise (hereafter SNR) requirements to obtain robust labels in this way are $SNR\sim 100$ per resolution element, especially if the labels are to include individual element abundances. 
Often, a post-calibration procedure is applied to bring the stellar labels derived by such a fitting pipeline in accord with external information of higher fidelity: for example with stellar labels from benchmark stars studied at high resolution or well characterised open and globular cluster stars \citep[e.g.,][]{Meszaros2013, Kord2013, Jofre2014}. These calibration stars are also used by surveys to provide a reasonable estimate of their label accuracy. 
In practice, different surveys or different pipelines end up delivering labels with different calibrations,
causing their stellar parameters or their abundances to be on slightly different scales.
This complicates inter-survey comparisons and constitutes a major challenge of the era of such large datasets. 

In this paper we propose and lay out a data-driven approach to deriving stellar labels from stellar spectra in the context of large spectroscopic surveys,
which we dub ``\tc''\footnote{The name \tc\ is inspired by the astronomer Annie Jump Cannon,
who was the pioneer in producing stellar classifications without any input of physical models!}.
The main practical strengths of \tc\ are that it requires no physical model of the spectra, it is enormously fast, it can obtain labels of comparable accuracy to that quoted in current physics-based approaches
 but at far lower SNR, and it offers a consistent way to cross-calibrate surveys. 
To achieve this, \tc\ relies on the existence of a subset of objects within a survey
(\textit{reference objects}) for which the stellar labels are known and cover label space sufficiently.

In this context, the term ``\emph{labels}'' refers to the pieces of information
that characterize and determine a stellar spectrum; these labels are commonly and sensibly split into \emph{stellar parameters} and \emph{element abundances}, although in the context of \tc\ it makes sense to treat them on a par. 
In most cases, it suffices to think of the labels as \teff , \logg, and the element abundances \xfe, athough stellar rotation, micro-turbulence, age, and so forth can also be thought of as labels.
It is central to the approach we lay out here that objects with the same labels have (nearly) identical spectra and that spectra vary smoothly with label changes. 
This must be true, if the set of labels is comprehensive enough so that it fully specifies the star; but if the labels are (for example) only \teff, \logg\ and \feh\ then this is an approximation. 
These three labels are typically described as stellar parameters and are by far the most important to describe the overall behaviour of the spectral flux of red giant stars.

There are fundamentally two steps in \tc. 
The first step, or \textit{training step}, is to create from the spectra of the reference objects a very flexible generative model (with $\sim 80,000$ parameters) that describes a probability density function (pdf) for the flux at every pixel in the continuum-normalized spectrum as a function of the labels.
The second step, or \textit{test step}, assumes that this same generative model holds for all the other objects in the survey (dubbed \textit{survey objects}). 
Then, the spectra of the survey objects and the generative model from the reference objects
allow us to solve for---or infer---the labels of the survey objects. 
Taken together the training step and the test step effect a \textit{label transfer},
transferring the known labels in the reference objects to the survey objects.

To make such an approach straightforward, we must assume that the reference objects and the survey objects were observed with an identical instrumental set-up, a condition well-satisfied with the large surveys listed above. 
We take the generative model for the continuum-normalized flux at each of $N_\pix$ pixels to be a polynomial function of all the labels, and hence the model is defined by its $N_\pix$ sets of polynomial coefficients. In practice, there may be different circumstances that make stars suitable reference objects. 
They may be members of star clusters: there, external data and the fact that clusters are in good approximation single stellar populations (which have to fall onto an isochrone) lend credibility to their stellar labels. 
Alternatively, reference objects could be stars for which labels have been derived separately from spectra of particularly high SNR, or at other ``easier'' or more extensive wavelength regimes (for example, in the optical as opposed to the infrared).  Finally, they may be subsets of stars for which other approaches to get stellar parameters (for example, astroseismology) provide accurate stellar labels.

\tc\ is a \emph{generative model} of the observed spectra; that is, it constructs, as a function of labels, a probability density function (pdf) for the observed flux as a function of wavelength. 
In many real cases the training data will be much higher in SNR than the test data (standard stars tend to be bright and well observed) and with \tc\ it is possible to transfer the labels from high SNR reference objects to lower SNR survey objects with high fidelity. In what follows, we show that \tc\ behaves very well as the SNR (or observing time) is decreased.

In this paper we use the \apogee\ survey as the sole example. 
However, \tc\ can be applied to any stellar survey.  

Our most basic implementation of \tc\ that we present includes only three labels, but this can easily be extended to additional labels  (for example, \alphafe, \xfe) and also more comprehensive models (for example, Gaussian processes). 
Additionally, as we are using the information in every pixel, this methodology is effective at determining labels at lower SNR than minimisation techniques.

\tc\ is similar to the MATrix Inversion for Spectral SythEsis (\matisse) and  University of Lyon Spectroscopic analysis Software (\textit{ULySS}) procedure for derivation of stellar parameters \citep{RB2006, Koleva2009} in that it uses the full spectrum
(and not just a line list) for label determination. 
However, \matisse\ employs a large grid of synthetic spectra
and is thus limited in all the ways that physics-based methods are limited. The method outlined in \citet{fio2007} proposes the possibility of using real data as a model but implements a different, principle component analysis technique to estimate stellar labels. Empirical stellar libraries have been used previously as a reference set of spectra, including with \textit{ULySS} \citep[e.g.][]{Wu2011, Prugniel2011} and also by \citet{Soubiran1998}, in order to determine stellar labels directly from observed spectra. A big part of why \tc\ is successful at lower SNR is that it uses all of the pixels in the data.

We have adopted a bottom-up approach for \tc, starting with the most basic implementation and successively adding complexity to the generative model to determine the least complex implementation that works. The aim of this paper is not to explore all possible models that may work for this approach, nor to converge on an optimal model, but to use the simplest model that validates the underlying methodology as successful. With additional complexity, for example partial labels on the reference objects and adding errors to the labels of these objects, it may be preferential to adopt a different form of model entirely, such as a Gaussian Process, considered in the Discussion Section. One key advantage of the simple model we adopt is in its relative simplicity, which makes \tc\ computationally trivial to run to return stellar labels for large datasets.

In laying out the methodology of this approach we firstly describe the \apogee\ dataset and the way we process the data for both reference objects (542 stars) and survey objects ($\sim$ 55,000 stars from DR10). 
We then describe perhaps the simplest implementation of label-transfer possible, using a first-order linear model. We found this first-order model to be insufficiently flexible to describe the labels of the stars and extended our model to quadratic form, which satisfactorily describes the label-space of the training data.
The success of this model is demonstrated by running \tc\ through the DR10 data available through the SDSS-3 data server, the results for which we provide in an online machine-readable table. 

\section{Data}\label{sec:Data}
\tc\ expects (in its simplest form, presented here)
all spectra---for reference and survey objects---to be continuum-normalized in a consistent way,
and sampled on a consistent rest-frame wavelength grid, with the same line-spread function.
It also assumes that the flux variance, from photon noise and other sources, is known at each spectral pixel of each spectrum.
In principle, \tc, as described below, is applicable to any large, homogeneous spectroscopic data set
meeting these criteria.
Here, we use the \apogee\ DR10 data (Majewski et al., 2015 (in prep)) to illustrate and showcase \tc.
Because all of the exposition of the method underlying \tc\ involves specificities of the data,
we spell out the characteristics of the \apogee\ data and our adjustments to it in \sectionname~\ref{sec:Apogee_as_worked_Example}. However, we stress that the approach is more widely applicable. 

\subsection{Specifics of the \apogee\ Data Set}
\label{sec:Apogee_as_worked_Example}

The data set used for this functional demonstration of \tc\ is that of the \apogee\ survey (Majewski et al. 2012, 2015 in prep). \apogee, part of the SDSS-III\footnote{\url{www.sdss.org}} (Eisenstein et al. 2011), is a high resolution (R $\sim$ 22,500), high signal to noise (SNR $\sim$ 100), H-band (15200-16900 \AA\footnote{Due to gaps between the instrument's three detectors, the spectra are divided into three pieces: $\sim$15150-15800~\AA, 15890-16430~\AA, and 16490-16900~\AA.}) spectroscopic survey of primarily red giant stars spanning the bulge, disk, and halo of the Milky Way \citep{Zaso2013}.  \apogee 's \aspcap\ pipeline provides the stellar labels for these stars, which include stellar parameters and multiple elemental abundances, in addition to numerous flags that warn of problems with the spectra or problems with the label determination for the spectra (or both).  This pipeline is based on $\chi^2$ fitting of the data to 1D LTE models for seven labels (\teff, \logg, \feh, [$\alpha$/Fe], [C/M], [N/M], and micro-turbulence; Garc\'{i}a~P\'{e}rez et al., 2015, in prep).

We use here spectra from the set of 55,000 stars that were released as part of the SDSS Data Release 10 (DR10) \citep{Ahn2014}), focusing on data from the \apstar\ and  \aspcapstar\ FITS files. The \apstar\ files include single-visit and combined spectra for a given star that are fully reduced, resampled, and shifted to the stellar rest frame.  The \aspcapstar\ files contain the combined spectrum for a given star that has also been pseudo-continuum normalized by the ASPCAP pipeline, along with the best-fitting synthetic spectrum and stellar labels.  The \apstar\ data, which are not pseudo-continuum-normalized by \apogee, enables us to evaluate the performance of \tc\ at lower SNR, by testing it on the individual visit spectra provided in these files. 

The pixel-by-pixel inverse variances are critical for all steps of \tc: continuum normalization, training step and test step. The error arrays of the uncertainty at each pixel in the spectra are provided by \apogee\ in their fits files. We adopt these vectors directly and additionally set any anomalous values in the spectra, with 0 flux or very high error values to a very large error value, for computational stability.  

Aside from photon noise, a number of other factors can contribute to the errors of any pixel in \apogee\ spectra: poor sky subtraction, cosmic rays, reduction induced errors, high persistence and other noise sources. In addition to the variance arrays, one can also use any bad pixel masks, where the
inverse variance and weighting of that pixel becomes $\sim$ 0. We find that adopting additional masking from the bad pixel masks degrades our results when using the combined \aspcapstar\ spectra. However, our results are improved for individual visit spectra in the \apstar\ files when pixels flagged in the bad pixel mask array provided by \apogee\ are rejected, by assigning the large weighing in the error on those pixels, for the individual visit spectra. We therefore only implement the bad pixel masks from \apogee\ for our tests on single visit spectra. 

The resampled, reduced and combined spectra are available for about 47,000 survey stars in 150 DR10 fields in the \aspcapstar\ files. There are a further 9000 star observed in commissioning mode only, which are available in the radial velocity combined but not continuum-normalized data format in the \apstar\ files.

We also apply \tc\ to the commissioning data and caution the reader about the fidelity of these results, as the Line Spread Function (LSF) of the commissioning data are different from the main survey and consequently different from the reference dataset of stars in the training step. Those objects are flagged as commissioning data (see Table 1). 
Typically, the line spread function for reference and all survey objects is the same within a given survey. As \tc\ calculates a separate spectral model for each survey it is applied to, survey homogeneity is sufficient; we do not need to know the actual LSF. This assumption breaks down if the survey stars are observed under a different instrumental setup to the reference objects, as is the case with commissioning data from \apogee. In this case, the LSF would have to be adjusted in a separate step, not introduced as a separate label; labels in the current context are strictly properties of the stars, not the experimental set-up.  

\subsection{Choosing Reference Objects for the Training Step}
\label{sec:ReferenceObjects}

For the training step in \tc\ we must choose a set of reference (or training) objects for which we have spectra from the survey under consideration and \emph{also} high-fidelity labels (that is, stellar parameters and element abundances that are deemed both accurate and precise).
The set of reference objects is critical, as the label transfer to the survey objects can only be as good as the quality of the reference label set. 
Also, as \tc\ may have to interpolate and extrapolate to new parts of label space as
it encounters new kinds of spectra among the survey objects, the quality
of the label transfer depends on the extent to which the reference objects
cover label space and the density with which they cover it.
The performance of a data-driven model like \tc\ will depend strongly on the size and quality of its training set of reference objects.

In practice, also for the \apogee\ data, one (but not only) good option
for a set of reference objects can be built from 
members of well-studied open and globular clusters that have been observed in the context of the survey \citep{Zaso2013, Meszaros2013}.
There is a variety of reasons why the stellar labels for cluster stars may be particularly accurate and robust.
For one, they could have their labels derived from independent, high resolution spectral analysis of these
stars, for example, from observations in a well understood portion of the optical wavelength region. The
labels are of course a property of the
star, and hence do not have to arise from the survey data at hand. They may have been derived from different data.

For the case of \apogee, we will use as reference objects 542 members of 19 globular and open clusters \citep{Meszaros2013}. These are the very same objects as used by the \apogee\ survey for their own calibration of the DR10 data release and represent the documented reference objects that are available. Some objects were removed from the full list available as their cluster memberships were incorrect. In their stellar labels they span the range of $3500<\teff<5300$~K, $0<\logg<5$ and $-2.5<\feh<0.45$. 

Exactly which stellar labels we adopt for these reference objects is critical to the subsequent output, and hence 
we discuss it in detail in \sectionname~\ref{sec:ApogeeRefLabels} .

\begin{figure}[h!]
\centering
    \includegraphics[scale=0.31]{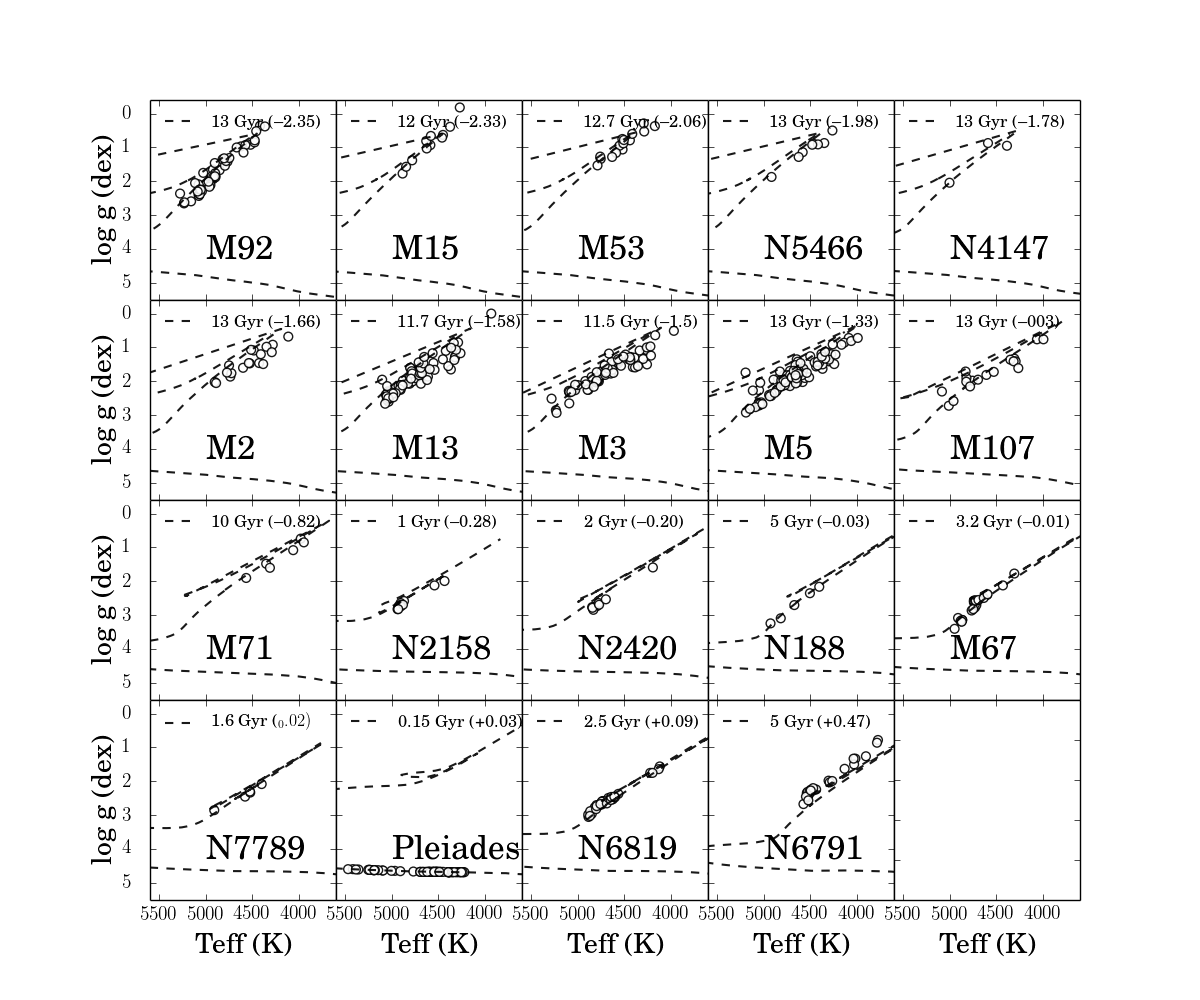}
\caption{\aspcap-corrected DR10 labels  for the training step in \teff-\logg\ plane for 542 stars in the 19 clusters for which parameters are provided by \apogee\ \citep{Meszaros2013}. The age and \feh\ of the isochrones (in parentheses) is shown in each sub-panel. All labels adopted from the \aspcap-corrected values of DR10 except for the Pleiades. }
\label{fig:trainingaspcap}
\end{figure}

\begin{figure}[h!]
\centering
  \includegraphics[scale=0.31]{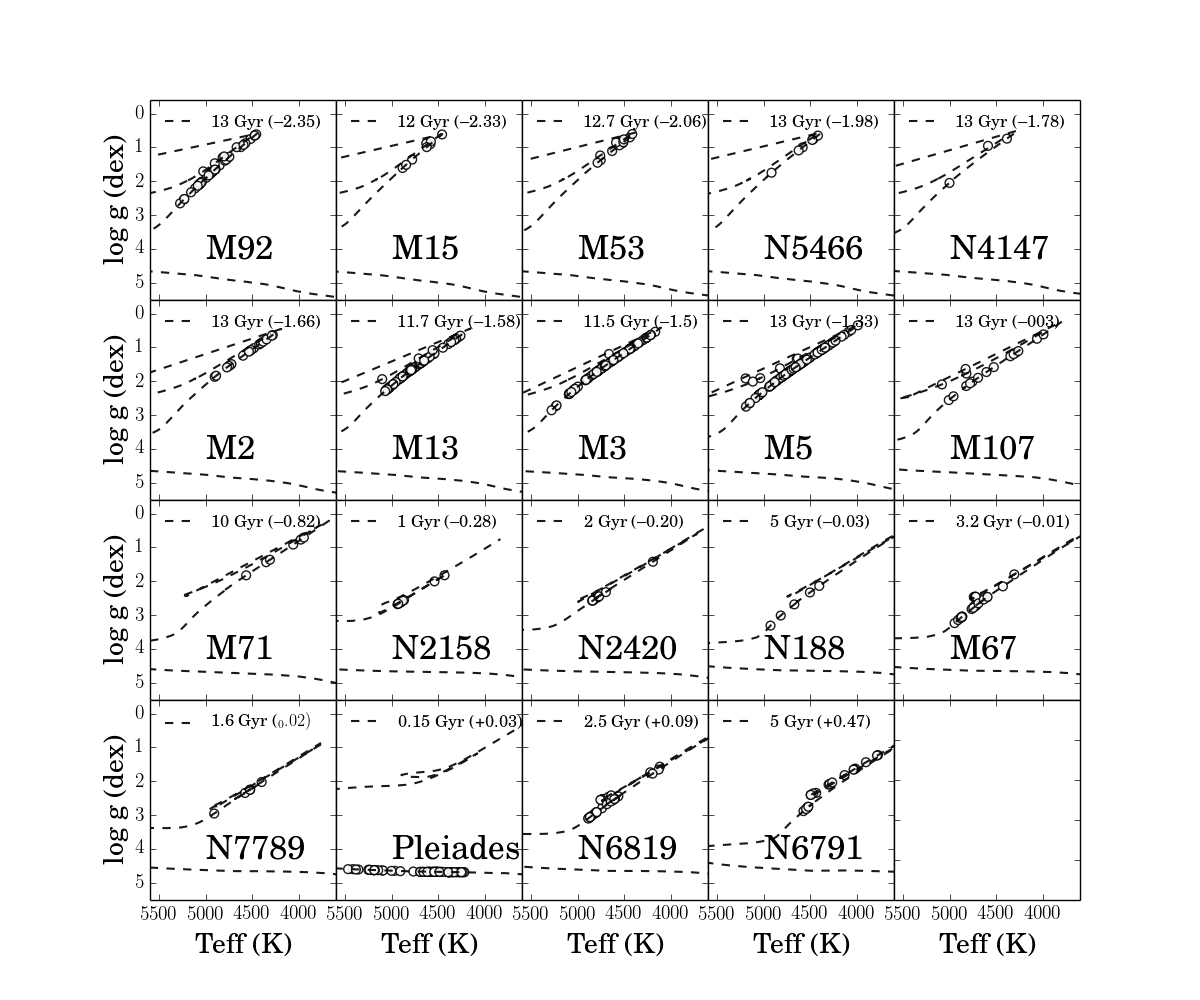}
\caption{Stellar labels for all reference objects as is \figurename~\ref{fig:trainingaspcap}, except that the \logg\ values have been adjusted from the \aspcap-corrected value to exactly match the isochrone, we refer to this set of labels as  ``isochrone-corrected'' labels to differentiate them from the correction in \figurename~\ref{fig:trainingaspcap}.  }
\label{fig:trainingisochrone}
\end{figure}

Another reason why cluster members make for good reference objects is because we can expect their stellar parameters to fall onto a single isochrone and to have near-identical abundances (at least for open clusters). This provides additional constraints on the labels.
We exploit that expectation in the case of \apogee\ and define ``Isochrone-corrected labels'', where we use Padova isochrones at the literature age and \feh\ of each cluster (see Figs. \ref{fig:trainingaspcap} \& \ref{fig:trainingisochrone}, and \sectionname~\ref{sec:ApogeeRefLabels}).

\subsection{Consistent Continuum-Normalization}\label{sec:ContNorm}

\tc\ operates on continuum-normalized spectra.
Continuum-normalization that is based on quantiles of the data (medians or 90-th percentiles or the like)
are very SNR dependent, for example, because pixels that are clearly \emph{not} continuum in high SNR
spectra are completely consistent with being continuum at lower SNR.
Therefore, to make \tc\ as independent of SNR as possible,
we base the continuum estimation on a pre-tabulated set of wavelength locations that we know
(iteratively, from running \tc\ itself, see Section~\label{sec:ApogeeContinuum}) are not strongly affected by absorption lines.

To initialize the continuum-pixel determination,
we define a preliminary pseudo-continuum-normalization by 
using polynomial fit to an upper quantile (for example, 80 or 90~percent) of the spectra, determined, for example, from a running median. For this pseudo continuum-normalized \apogee\ spectra we use a running quantile across 50~\AA~ of the spectra, taking the 90th percentile. This is effective, but SNR-dependent.
  
After a training step using spectra of reference objects that have been  normalized  by this pseudo-continuum,
 \tc\ can provide an improved identification of continuum regions in the spectrum: 
we take those pixels to be continuum that show nearly unity flux in the spectral model's baseline spectrum (see \sectionname~\ref{sec:spectralmodel}), and at the same time show almost no dependence in their normalized flux on the stellar labels.
That is, for the \apogee\ data, we can determine with \tc\ the `true' continuum, using the model derived from the pseudo-continuum-normalized spectra for the reference objects provided by \apogee, as described in \sectionname~\ref{sec:results}. This constitutes a data-driven method for finding continuum pixels, and we find it to have only a very small systematic dependence of the spectra on SNR (see \sectionname~\ref{sec:ApogeeContinuum}).

Given the continuum pixels, we implement a least-squares fitting to the \apogee\ spectra of a low-order Chebyshev polynomial, fitting only to the determined continuum pixels outlined in Section ~\ref{sec:ApogeeContinuum}. We treat each of the three chips separately, and find a 2nd-order Chebyshev polynomial to be sufficient to apply to the data provided by \apogee. We apply this normalisation to both \aspcapstar\ and \apstar\ files. Treating the three chips separately, we fit the polynomials over the wavelength regions of (i) 15150-15800~\AA, (ii) 15890 16430~\AA~ and (iii) 16490 - 16950~\AA. Fitting a polynomial has the disadvantage that they are poorly constrained at the edges of the data. An alternative implementation could use a more sophisticated sine or cosine function in place of a polynomial fit. 

\begin{figure}[h!]
  \includegraphics[width=\hsize]{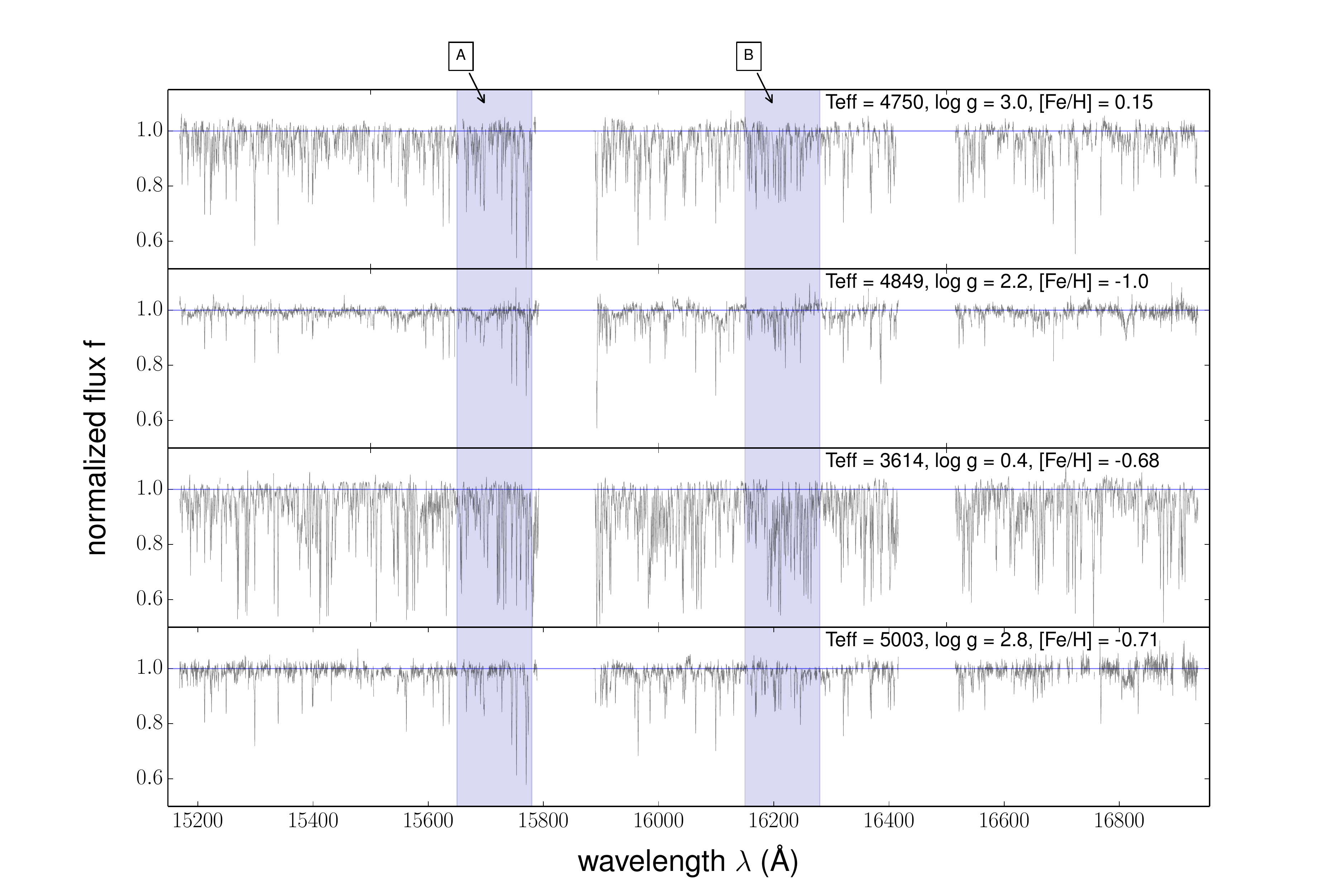}
\caption{Continuum-normalised spectra for stars across a range of stellar labels; at top, two stars of similar temperatures at different metallicities and at bottom, two stars of similar metallicities and different temperatures. The grey shaded regions A and B indicate \apogee\ sample wavelength regions used for subsequent \figurenames\ in the paper.}
\label{fig:norm}
\end{figure}

\figurename~\ref{fig:norm} shows an example of this iterated normalisation applied to survey spectra with different labels. To illustrate the result, \figurename~\ref{fig:norm} shows typical \apogee\ spectra and demonstrates how the spectra 
vary as a function of metallicity at a given temperature, and as a function of temperature at a given metallicity. 
For a clearer view of individual absorption line features, we use narrower regions marked in this \figurename, (A) and (B), for all subsequent examination of the spectral data.

\subsection{Labels for the Reference Objects in \apogee}
\label{sec:ApogeeRefLabels}

Which values to adopt for the labels of the reference objects used in \tc 's training step is a critical issue
in any survey. We discuss two options here for \apogee . 
First, we adopt the DR10 \textit{``\aspcap-corrected''} stellar parameters \citep{Meszaros2013} that are available for each of the reference objects as their labels, in order to place the output of \tc 's test step for the survey objects on the \apogee\ \aspcap\ scale (\figurename~\ref{fig:trainingaspcap}). ``\aspcap-corrected'' labels were not available for the cluster comprised of main sequence stars, the Pleiades cluster and for this cluster we made our own corrections in \teff\ and \logg\ and assumed a single literature value for the \feh\ label, as described below. 
The reference set of stars we use is the very same stars used by \apogee\ to post-calibrate the output of \aspcap\ to a physical stellar parameter scale \citep{Meszaros2013}.
Adopting the \aspcap-corrected labels provided and documented by \apogee\ has the important advantage 
that we can test exactly how well we can reproduce the results from \apogee\ for the survey stars via label-transfer from only 542 stars. A limitation of this reference set of objects is that main sequence stars are not well sampled, which, as we discuss in Section 5, limits our ability to determine labels for these stars at the test step. 

The corrections to the labels made by \apogee\ described in \citet{Meszaros2013} are based on the cluster data and applied to the immediate output of the \aspcap\ pipeline that arose from comparisons to a library of stellar models.
Temperature corrections are determined by comparing the infrared flux temperatures of the stars \citep{gonzalez2009}, \logg\ corrections are from the offset between \aspcap\ results and Kepler astroseismic results for stars in common and \feh\ corrections are from the difference between the \aspcap\ and  the literature value of each cluster.  
The \apogee\ corrections determined in \citet{Meszaros2013} are valid only for stars with \logg\ $<$ 3.5 and are not implemented for the dwarfs.  We adopt the \aspcap-corrected \mh\ values for these clusters and these are corrected to the \feh\ of the clusters and so we adopt this label as an \feh\ (that is, this label from \apogee\ therefore, does not explicitly use \feh\ lines, but is derived from an \feh\ correction). 
The analysis in \citet{Meszaros2013} is restricted not only to giants but also stars with SNR $>$ 70, determined to be the minimum SNR for reliable stellar parameters by \apogee.

These corrections implemented by \apogee\ in \teff, \logg\ and  \feh\ place the giants in the cluster stars on or near the iscohrones (see Figures 7 and 8 in Meszaros et al., 2013).  
As there are no \aspcap\ corrections implemented for the 65 main sequence stars among the reference objects that we use, 
we instead determine temperatures for these dwarfs, which are all in the Pleiades, using the same correction method 
as in \citet{Meszaros2013}. 
We determine the infrared flux temperature for the stars from \citet{gonzalez2009} and apply a correction to 
the \aspcap\ output based on the offset in the temperature scales. For the dwarf stars in the Pleiades, we find the following relation:
 $T_{\mbox{corrected}}$= 0.855*T$_{\mbox{\textit{\aspcap}}}$ + 1206.7.

We do not attempt an individual metallicity correction for each dwarf star in the Pleiades but rather set all \feh\ of the dwarf spectra to \feh\ = 0.03 \citep{barrado2001}.
Tests on the input labels to \tc\ demonstrate that there is only a small degradation of the results caused by adopting a single \feh\ for every cluster star for the literature value of the cluster, instead of individual \aspcap-corrected \feh\ values for the stars.
To determine the \logg\ for these Pleiades main sequence stars, we shift the stars vertically to their nearest positions on an appropriate age-metallicity Padova isochrone of 150~Myr at $\feh = 0.03$ \citep{girardi2000}. 
Due to the high differential reddening to the Pleiades, and the subsequent large temperature errors using the IR flux method that result from this, we only selected the 65 from a total of 72 Pleiades dwarfs, eliminating those with high extinction of SFD (corrected) E(J-K) $>$ 0.30 \citep{Schlafly2011}.

Adopting the input labels from the \aspcap-corrected parameters determined from calibrations to literature cluster values also transfers the errors from the \aspcap\ pipeline: of $<$ 150K in \teff,  $<$ 0.2 dex in \logg\ and $<$ 0.1 dex in \feh.   
The uncertainties on the input labels will be included as an input parameter of the labels in a future development stage of \tc. Inclusion of uncertainties may be particularly relevant when introducing multiple labels of individual elements. 

For a comparative analysis to the ``\aspcap-corrected'' labels (Figure \ref{fig:trainingisochrone}), we adopt a \logg\ label for all of the training stars not from the Kepler scale, but rather from the best vertical fits to the isochrone for the ages and metallicities for the clusters from the literature (with the temperatures fixed).
We call these the ``Isochrone-corrected'' labels, where we use Padova isochrones at the age and \feh\ of each cluster. 

\section{\tc 's Training Step: Making a Generative Model}
\label{sec:spectralmodel}

We now lay out the spectral model, whose parameters are determined 
from the spectra and stellar labels of the reference objects in the training step.
Such a generative model is based on two basic notions: first, that the continuum-normalized spectra of
stars with identical labels look near-identical at every pixel, save for the observational errors
and some intrinsic scatter. This must be true if the set of labels is exhaustive. 
In practice, that is an approximation, as for example, the spectra of stars with identical \teff , \logg \ and \feh\ may differ, 
as these stars have different \alphafe , age or rotation. Second, we presume that the expected flux at every pixel changes continuously
with changes in the labels.
Importantly, the model is a probabilistic generative model that produces,
for every object spectrum at every wavelength,
a pdf for the flux, with an expectation value (mean) and a variance.

We presume there are $N_\rfn$ reference objects $n$, each of which has
a continuum-normalized flux measurement $f_{n\lambda}$ at wavelength
$\lambda$. Each of the training spectra (of index) $n$ has $K$ labels $\starlabel_{nk}$, each of which
is (for now) presumed to have negligible uncertainty and contained (possibly with transformations; given below)
within a label vector $\starlabelvec_n$.

We then presume that for any star, $n$ at any pixel, $\lambda$  the flux $f_{n\lambda}$  can be described as some smooth function of the star's labels $\starlabel_{nk}$
($\teff,\logg,\feh,\cdots$).
The observations $f_{n\lambda}$ will differ from such a model by the observational noise (from all relevant sources), $\noise$. But even for perfect measurements we presume that there will be
deviations from the above approximate model for the true flux, characterized by a scatter $\scatter$,
which is a property of any particular pixel; we will subsume $\scatter$ under the noise.

Generally, we take a spectral model to be characterized by a coefficient vector $\set{\theta}_\lambda$
that allows to predict the flux at every pixel $f_{n\lambda}$ for a given label vector 
$\starlabelvec_n$:
\begin{eqnarray}
f_{n\lambda} &=&
g(\starlabelvec_n |  \set{\theta}_\lambda) + \mbox{noise}
\label{eq:specmodel}\quad 
\end{eqnarray}
As a specific, but still flexible functional form for the spectral model we presume that it can be written as
a linear function of some vector $\starlabelvec_n$ built from the labels: 
\begin{eqnarray}
f_{n\lambda} &=&
\set{\theta}_\lambda^T \cdot \starlabelvec_n + \mbox{noise}
\label{eq:linearmodel}\quad
\end{eqnarray}
where $\set{\theta}_\lambda$ is the set of spectral model coefficients at each $\lambda$. Each element of $\starlabelvec_n$ can be some (possibly complicated) function of the full set of $K$ labels, $\starlabelvec_n$, which
results in the flexibility of this model. The noise is an \textit{rms} combination of the associated uncertainty variance
$\sigma_{n\lambda}^2$ of each of the pixels of the flux from finite photon counts and instrumental effects and the intrinsic variance or scatter of the model at each wavelength of the fit, $s_\lambda^2$.
This model assumes that the noise model is
$\mbox{noise} = [s_\lambda^2+ \sigma_{n\lambda}^2]\,\xi_{n\lambda}$,
where each $\xi_{n\lambda}$ is a Gaussian random number with zero mean and unit
variance.

The simplest spectral model is that in which the label vector $\starlabelvec_n$ is
linear in the labels, that is, in the vector of the individual labels themselves:
\begin{eqnarray}
\starlabelvec_n &\equiv& [1,
                           \starlabel_{n1} - \mean{\starlabel_1},
                           \starlabel_{n2} - \mean{\starlabel_2},
                           \cdots,
                           \starlabel_{nK} - \mean{\starlabel_K}]
\label{eq:linear}\quad,
\end{eqnarray}
where the first element ``1'' will permit a linear offset in the fitting.
The $\mean{\starlabel_k}$ are offsets (possibly means of the training data) to
keep the model ``pivoting'' around a reasonable point in label space.
This model leads to the single-pixel log-likelihood function 
\begin{eqnarray}
\ln p(f_{n\lambda}\given\set{\theta}^T_\lambda, \starlabelvec_n, s_\lambda^2) &=&
 -\frac{1}{2}\,\frac{[f_{n\lambda} - \set{\theta}^T_\lambda \cdot \starlabelvec_n]^2}{s_\lambda^2 + \sigma_{n\lambda}^2}
 -\frac{1}{2}\,\ln(s_\lambda^2 + \sigma_{n\lambda}^2)
\label{eq:like}\quad.
\end{eqnarray}

The vector $\set{f}_\lambda$ is the set of spectral flux values for
all $N$ objects at the one wavelength $\lambda$. This is a likelihood function for the labels and parameters:  It provides a probability density function evaluation at the data given settings of all the labels and parameters.
We can set the coefficients $[\set{\theta}_\lambda,s_\lambda^2]$ either by
optimizing the likelihood (\ref{eq:like}) over all reference objects or by applying priors and
performing some form of probabilistic inference (with, say, Markov
Chain Monte Carlo techniques).
Here we will optimize for now, which can be done separately for each pixel $\lambda$, where
we are treating the spectral model coefficients $\set{\theta}_\lambda$ and the scatter $s_\lambda^2$ as free parameters, and the
labels in the label vector $\starlabelvec_n$, $\starlabel_{nk}$ as fixed:

Then, in the training step of \tc\ we exploit the fact that we know the $f_{n\lambda}$
and the $\starlabelvec_n$, which permits to solve for the coefficients and the scatter of the spectral model:
\begin{eqnarray}
\set{\theta}_\lambda,s_\lambda \leftarrow \substack{\mbox{argmax}\\{\set{\theta}_\lambda}, s_\lambda}
\sum_{n=1}^N \ln p(f_{n\lambda}\given\set{\theta}^T_\lambda, \starlabelvec_n, s_\lambda^2)
\label{eq:trainingstep}
\end{eqnarray}
The linear-in-labels form (\ref{eq:linear}) has a number of useful properties.
The coefficient vector $\theta_{\lambda 0}$ has a simple interpretation;
it is the ``baseline spectrum" of the spectral model. 
The next coefficient vectors,  $\theta_{\lambda k}$, linear in \teff , \logg ~and \feh,
describe the lowest-order dependence of the spectrum on these labels.
In practical terms, the optimization of the model parameters $\theta_{\lambda k}$, at fixed scatter
$s_\lambda^2$ is a pure linear-algebra operation (weighted least
squares); simultaneous optimization of all the parameters
$[\set{\theta}_\lambda,s_\lambda^2]$ is only nonlinear in the $s_\lambda^2$
parameter.

The (perhaps) second-simplest spectral model is that in which the
vector $\starlabelvec_n$ is quadratic in the labels: so this label vector is described as:
\begin{eqnarray}
\starlabelvec_n &\equiv& \begin{array}{l}[1,
                          \starlabel_{n1} - \bar{\starlabel_1},
                          \starlabel_{n2} - \bar{\starlabel_2},
                          \cdots,
                          \starlabel_{nK} - \bar{\starlabel_K},\\
                          (\starlabel_{n1} - \bar{\starlabel_1})\,(\starlabel_{n1} - \bar{\starlabel_1}),
                          (\starlabel_{n1} - \bar{\starlabel_1})\,(\starlabel_{n2} - \bar{\starlabel_2}),
                          \cdots,\\
                          (\starlabel_{nK} - \bar{\starlabel_K})\,(\starlabel_{nK} - \bar{\starlabel_K})]\quad ,
\end{array}
\label{eq:quadinlabels}
\end{eqnarray}
where the quadratic terms contain all possible products exactly once.

For the training step of \tc , this quadratic-in-labels form of the spectral model (\ref{eq:quadinlabels}) is similar to a the linear-in-labels form (\ref{eq:linear}) in a number
of ways.
It is still the case that optimization of the model, at fixed scatter
$s_\lambda^2$ is a pure linear-algebra operation (weighted least
squares), except that $\starlabelvec_n$ has become longer for a given number of labels. 
However, the test step on the survey (described in the next Section) of the quadratic-in-labels form
 will no longer be simple; it will require non-linear
optimization to estimate the labels.

The coefficients $\theta_{\lambda 0}$ can still be seen as an estimate of the
\emph{baseline spectrum} (provided that the offsets $\mean{\starlabel_k}$ are the
mean tag values); the first-order coefficients $\theta_{\lambda k}$ can still
be seen as first derivatives of the expected spectrum with respect to
each of the $k$ labels, but now evaluated at the baseline spectrum; the
second-order coefficients $\theta_{\lambda kk'}$ can now be seen as mean
second derivatives of the expected spectrum with respect to pairs of
labels $k$ and $k'$.

\section{\tc's Test Step: Labeling Survey Spectra}
\label{sec:paramestimate}

In the previous Section, we trained or fit the parameters of
a data-driven probabilistic generative model for stellar spectra from the reference objects
serving as training data.
This model has the property that, given labels (and noise variance estimates), it produces a
pdf for the continuum-normalized flux, that includes both observational and intrinsic
scatter.
In this \sectionname, we are going to solve the inverse problem:
we have spectra, but we don't have labels for them.
In this case, we will use inference and the just determined spectral model
to obtain labels for the untagged survey
spectra, which we also refer to as the ``test data'' in what follows. 

In the test data there will be $M$ spectra $m$, each of which---as in
the training data---has a continuum-normalized flux measurement
$f_{m\lambda}$ at each wavelength $\lambda$, and an
associated observational uncertainty variance $\sigma_{m\lambda}^2$.
Just as in the training step, we consider the same likelihood function given in
Equation~(\ref{eq:like}). But now we view it as a function of the \emph{labels},
instead of the function parameters $\set{\theta}_\lambda$ and
scatter $s_\lambda^2$.

In the test step of \tc\ we use the spectral model coefficients and scatter,
($\set{\theta}_\lambda,\ s_\lambda^2$), to be exactly those that were determined in the training step.
We then take the entire $N_\pix$ spectrum of survey star $m$, $f_{m\lambda}$ and optimize for the labels of that star:
\begin{eqnarray}
\left\{\starlabel_{mk}\right\} \leftarrow \substack{\mbox{argmax}\\{\left\{\starlabel_{mk}\right\}}}
\sum_{\lambda=1}^{N_\pix}
\ln p(f_{m\lambda}\given\set{\theta}^T_\lambda, \starlabelvec_m, s_\lambda^2)
\label{eq:teststep}\quad .
\end{eqnarray}
The labels $\starlabel_{mk}$ for each survey star $m$ can be obtained either by maximizing
the likelihood function, or else by applying priors
and performing probabilistic inference.
Again, we will optimize here. Our optimization is not convex in general, but in practice it is insensitive to initialization.
The right-hand sides of the training step (\ref{eq:trainingstep}) and test step (\ref{eq:teststep}) look formally quite analogous.
But in the test step we optimize over the labels, considering all pixels of one survey object at a time. In contrast,
in the training step, we optimize over the spectral model coefficients and scatter, considering all reference objects
at one pixel at a time.

When we use the simple linear-in-labels form (\ref{eq:linear}) for the
mean model, the optimization to obtain maximum-likelihood labels
(given parameters $[\set{\theta}_\lambda, s_\lambda^2]$) is simple linear
least-square fitting.
This optimization is obtained by straightforward linear algebra on the
spectral pixels $f_{m\lambda}$, and standard frequentist confidence
intervals can be obtained similarly.
When we use the quadratic-in-labels form (\ref{eq:quadinlabels}) for the
spectral, there is no simple linear-algebra operation that
optimizes the likelihood. 
Instead an optimization function is used, the python curve$\_$fit routine, which uses a non-linear least squares fit to fit the function to the data. 

We have described how we construct a spectral model from the reference objects in the training step and then 
estimate stellar labels for survey stars with that model in the test step. 
We now present in \sectionname~\ref{sec:results} the results of implementing our model for all \apogee\ data, where we applied a quadratic model: linear in the coefficients and non-linear in the label-inference.  
For the quadratic model we then show this applied to the DR10 data, including at lower SNR, and investigate different input training labels. 

\section{Results with \apogee\ Data}
\label{sec:results}

We now present the results for applying \tc\ to \apogee\ data. 

To apply \tc\ to \apogee\ data, we first train the quadratic model in Equation~\ref{eq:quadinlabels} using the reference data and three labels chosen as described in Section~\ref{sec:ApogeeRefLabels}. We then apply this model to all of the DR10 continuum-normalized data, using continuum-normalized \aspcapstar\ spectra described in Section~ \ref{sec:ApogeeContinuum}. We use a leave-one-out cross-validation test to explore which complexity the spectral model must have and how comprehensive the set of reference objects should be. We then proceed with the same set of reference objects in the label transfer to the entire DR10 in the test step. In particular, we also apply the test step to spectra from individual \apogee\ visits that have far shorter exposure times, and hence lower SNR than the co-added spectra in DR10, in order to explore and illustrate how well \tc\ does at modest SNR (with appropriate continuum fitting).

\subsection{The Choice of the Spectral Model Complexity}
\label{sec:ModelComplexity} 

To evaluate \tc 's label-transfer we have to settle on a suitable functional form for the spectral model (\ref{eq:specmodel}).
To start, one could consider picking the simplest -- namely linear-in-label -- spectral model, comprised of only four coefficients at every pixel (\ref{eq:linear}).
However, through take-one-out tests on the set of reference objects (see \sectionname~\ref{sec:take-one-out}), we found that this simple linear model was too inflexible to describe the spectral flux dependence on the labels.
As a consequence, the labels that emerged from the test step applied to the reference objects showed large and systematic deviations compared to ``known" input label values,
especially at the extremes of the labels' ranges. 

This is perhaps not surprising, as absorption features, particularly strong lines, are known to vary non-linearly as a function of stellar labels. If one were to insist on a label transfer with a first-order, spectral model, the systematic discrepancies could presumably be reduced by selecting only weak-line regions, but at the severe price of leaving much of the spectral range unexploited. Therefore, we have not pursued the linear-in-labels \textit{Ansatz} for the spectral model.

The next simplest spectral model, the quadratic-in-labels case,
 presumes that the continuum-normalized flux is a general second-order polynomial of the stellar labels, $f_{n\lambda} =
\set{\theta}_\lambda^T \cdot \starlabelvec_n + \mbox{noise}$ 
(\ref{eq:linearmodel}), 
but where $\set{\theta}_\lambda$ now contains 10 elements at every pixel.
For the case of the three labels $(\teff , \logg , \feh)$ the label vector $\starlabelvec_n$
becomes  
\begin{eqnarray}
\starlabelvec_n &\equiv&
[1, \teff, \logg, \feh, \teff^2, \teff\cdot\logg, \teff\cdot\feh, \logg^2, \logg\cdot\feh, \feh^2]
 \label{eq:quadinthreelabels}\quad.
\end{eqnarray}

We will use this quadratic-in-labels spectral model throughout the rest of the paper.  The exploration of higher-order polynomials for the spectral model at every pixel, or even a Gaussian process at every pixel, is beyond the scope of this paper. 
For any other application the complexity of the spectral model (for example, is a quadratic model good enough?) should be examined.
 
\subsection{Validation on Take-One-Out Stars from the Reference Objects}
\label{sec:take-one-out}

\begin{figure}[h!]
\centering
    \includegraphics[scale=0.45]{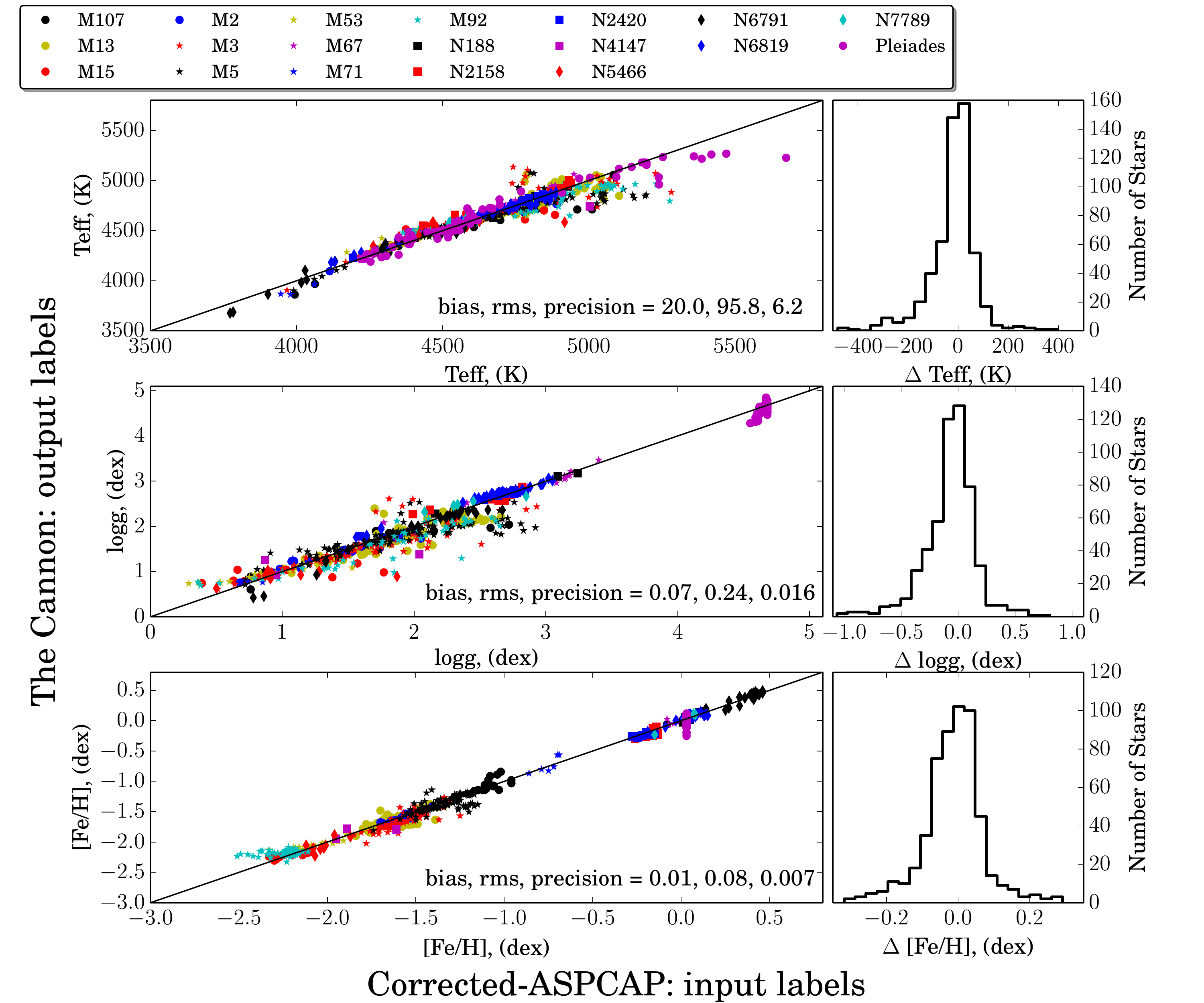}
\caption{The take-one-star-out cross-validation of the 542 stars in the training dataset using the quadratic model in Equation~(\ref{eq:quadinthreelabels}) and corresponding histograms at right, showing \tc\ output -- \apogee\ input labels.}
\label{fig:takeonestarout}
\end{figure}

As a first illustration of how well \tc\ works in practice, we perform a take-one-star out test on the set of reference objects.
For the take-one-star out test we train the spectral model iteratively on the spectra of all but one of the $N_\rfn$ (=542) 
reference objects, and then apply \tc 's test step to the spectrum of that remaining object. If we repeat this procedure $N_\rfn$ times, 
we have a first powerful test of how the result of this parameter transfer compares to the (known) labels for the reference objects.
 Here we only consider three labels $(\teff , \logg , \feh)$, and the results are shown in \figurename~\ref{fig:takeonestarout}.

This \figurename\ clearly shows how well \tc\ works, at least in the circumstance at hand.
\tc 's purely mathematical approach of label transfer estimates the stellar labels (at least) as well as the astrophysical \aspcap\ pipeline,
over the full label range of our the reference data. The \textit{rms} of the difference between the \aspcap\ and \tc\ values for the three labels are
95~K in \teff, 0.24 in in \logg, 0.08 in \feh, with biases $\Delta$ that are 3-7 times smaller.  
These variances inherently include some portion of the uncertainties on the input labels (from ASCPAP corrected values, 
of \teff\ $<$ 150 K, \logg\ $<$ 0.2 dex and \feh\ $<$ 0.1 dex \citep{Meszaros2013}).
The precision values stated in \figurename~\ref{fig:takeonestarout} are the formal uncertainties in the labels arising 
in the test step's optimization; for the SNR of the spectra in this take-one-out test, these errors are very small.
It is important to remember that the one left-out object and its spectrum are completely detached from the training step, 
except that they have the same experimental set-up and are likely drawn from a part of label space well-represented by the remaining reference objects.

There are a few outliers in \figurename~\ref{fig:takeonestarout}, cluster members of M3 in particular, that are offset in \teff\ and \logg\ space. 
The Pleiades cluster, which has only spectra for main sequence stars, shows the poorest determination in the \feh\ label. We assigned all its members a single \feh ~as reference labels, unlike the other reference objects, where we used their \aspcap -corrected labels from DR10.
The \textit{rms} is comparable to the estimated \apogee\ errors. The \logg\ label has the largest relative \textit{rms} in the \aspcap\--\tc\ comparison, larger than the \apogee\ uncertainty, suggesting an internal uncertainty of $<$ 0.1 dex in \logg\ determined by \tc.
If we adopt instead of \aspcap -corrected \logg\ labels the isochrone-corrected \logg 's (see \sectionname~\ref{sec:ApogeeRefLabels}), the \textit{rms} improves by 10\% in \teff\ and \logg.

The outlying stars in \figurename~\ref{fig:takeonestarout} may be due to an anomalous scale of the input labels of these stars compared to the other training data, or it may be a consequence of the model being too inflexible to properly describe how flux changes with labels across the parameter space of the training dataset. 
The temperature of the dwarfs is offset low at increasing temperature, compared to the input labels, so the model may be limited in describing the difference between dwarf and giant spectra. 
There is a flattening at the low metallicity end of the model in \feh\ in the output labels at \feh\ $<$ -2.2. However this value of \feh\ = --2.2 also corresponds to the literature value of this cluster, M5 \citep{Meszaros2013}. 
The lower metallicity of the \aspcap\ label may represent internal scatter in the \aspcap\ results.
The fact that \figurename~\ref{fig:takeonestarout} shows only very small systematic offsets and such tight scatter leads us to conclude that for the
current context the quadratic-in-labels spectral model is sufficient in the label transfer. 

Interestingly, an analogous take-one-cluster-out test significantly increases the scatter in the label transfer, 
increasing the \textit{rms} differences to $<$ 150 K in \teff, $<$ 0.4 dex in \logg\ and $<$ 0.12 dex in \feh.
This indicates that our training set is sufficiently small that each cluster matters for a good label transfer. 
One particular case in the \apogee\ context is the Pleiades cluster: it is the \textit{only} cluster for which dwarf stars have been observed and hence we can draw reference labels for main sequence stars.

\begin{figure}[h!]
\centering
    \includegraphics[width=\hsize]{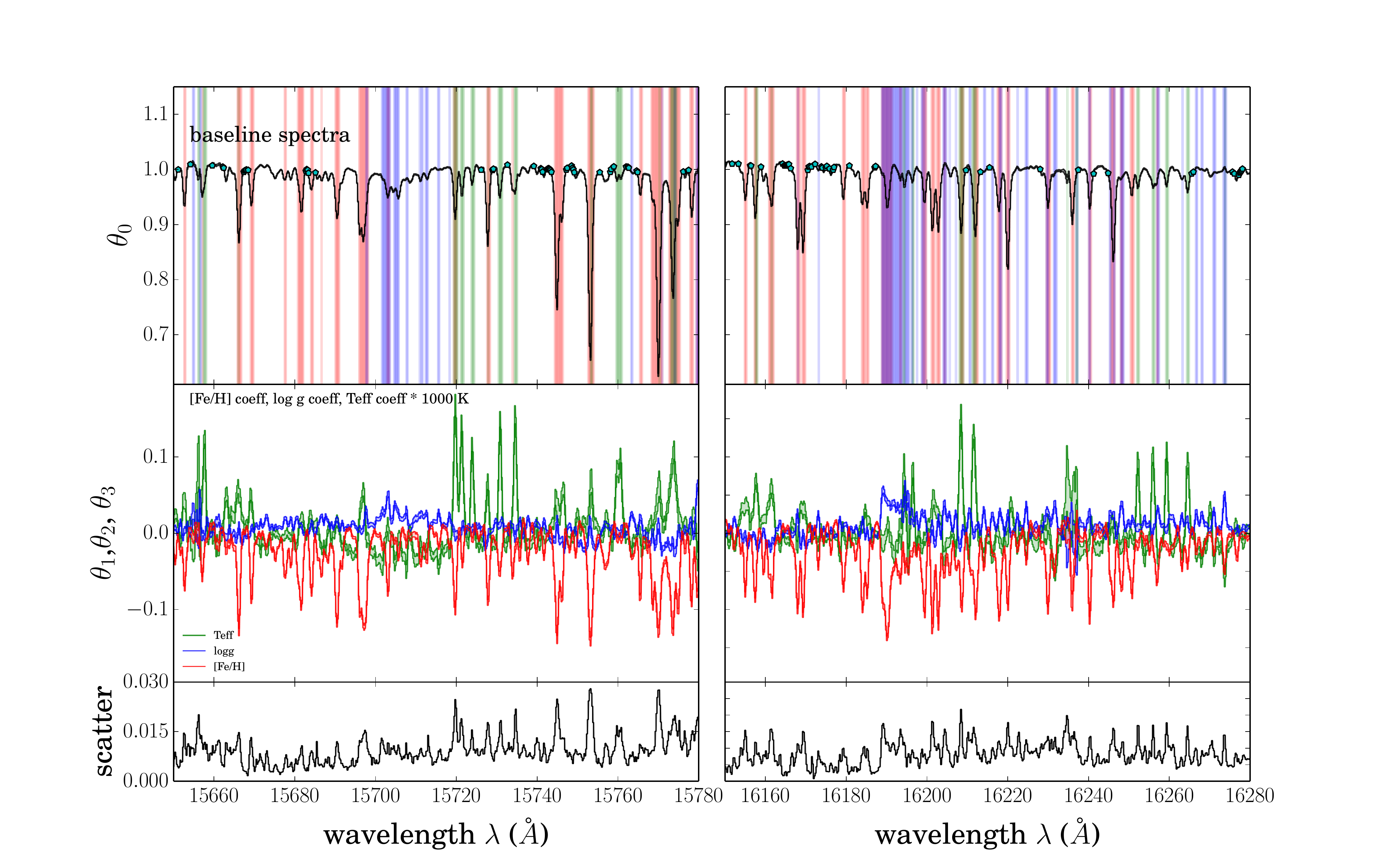}
  \caption{The first-order coefficients and scatter across the sample regions of the spectra from \figurename~\ref{fig:norm}, A and B. Top panel: the baseline spectra representing the first coefficient from the set of reference spectra; middle panel: the next three coefficients ($\theta_1$, $\theta_2$, $\theta_3$),  which correspond to the labels ($\teff, \logg, \feh$); bottom panel: the scatter of the fit with a tenfold expanded vertical scale. The red, blue and green areas in the top panel encompass the wavelength regions with the 5\% highest (absolute value) coefficients for the \feh, \logg\  and \teff\ labels respectively. The \teff\ coefficient has been multiplied by a factor of 1000 simply to show this coefficient on a similar scale to the other coefficients. This indicates where the flux in these spectrum is particularly sensitive to the labels.  Note that the \feh\ label is dominant in the contribution level and from the top panel it is clear that there is significant covariance between the labels and there are only a few regions of \logg\ sensitivity. The filled dots in the baseline spectrum in the top penal indicate the wavelengths at which the dependencies on all labels are weak, which we operatively identify as continuum pixels (see \sectionname~\ref{sec:ApogeeContinuum}).}
\label{fig:coeffs}
\end{figure}

We now turn to illustrating where the information that led to the accurate label transfer (\figurename~\ref{fig:takeonestarout}) came from in the spectra.
\figurename~\ref{fig:coeffs} shows -- across the narrow regions (A) and (B) of the spectra, marked in \figurename~\ref{fig:norm} -- the first coefficient vectors $\theta_{0,1,2,3}$ of the spectral model (those linear in the three labels), which were fit for in the training step for the quadratic-in-labels model in Equation~(\ref{eq:quadinlabels}).

The top panel of \figurename~\ref{fig:coeffs} shows the zeroth order-coefficient vector $\theta_0$, or the baseline spectrum, of the model. 
The mid panel shows the coefficients that are simply linear in \teff, \logg\ and \feh.
In the top panel of \figurename~\ref{fig:coeffs}, the red, blue and green shaded wavelength regions with the 5\% 
highest coefficient values $|\theta_{1,2,3}|$ in the \feh, \logg\ and \teff\ labels respectively. 
These regions indicate where the spectra's flux levels strongly vary with these labels.
This also highlights that different parts of the spectrum depend differently on the labels. Note there are many regions where the \feh\ label dominates in contribution to the flux.
For the first label vector for example in the middle panel of \figurename~\ref{fig:coeffs}, 
there is typically asymmetry for a given absorption feature, in the flux and the labels. 
There are very few regions where the flux is a function of only one of the labels, and pixels are typically co-variant. 
(that is, the same pixel will have a higher flux at both lower \teff\ and higher \feh). 
This simply reflects well-known co-variances between, for example, temperature and \feh .
The strongest \logg\ dependence is typically associated with weak lines including the wings of the 
feature and the \feh\ label, with strong lines, particularly the depth of the line. 

The bottom panel of \figurename~\ref{fig:coeffs} shows the scatter vector of the spectral model, 
indicating the dispersion of the flux of the training data around the best-fit spectral model at each pixel. 
The scatter is small and this indicates that our model is a good representation of the data. 
However, the scatter is highest where the most information in the spectra are contained. 
This implies that either our quadratic-in-labels spectral model is still somewhat too restricted, or that the labels of our training dataset are imperfect or incomplete 
(for example, lacking $[\alpha / Fe]$ as a label), or a combination of these effects. 
From the coefficients of an initial fit of this spectral models (see, for example, the middle panel of \figurename~\ref{fig:coeffs}), 
the continuum pixels have been determined following \sectionname~\ref{sec:ContNorm}. 
These are marked in the cyan dots in the top panel of the \figurename, and are used for an iterated, 
consistent continuum-normalization for all spectra, both of the reference and of the survey objects.

To demonstrate the fit of the model to the \apogee\ spectra at test time, using the continuum normalised test stars shown in Figure 3, we plot the best fit model and the corresponding spectra for the regions A and B highlighted in this Figure. This is shown in Figure 6, which demonstrates that the model is an excellent fit to the data: the three labels used to describe the flux are sufficient.

\begin{figure}[!h]
\centering
 \includegraphics[scale=0.4]{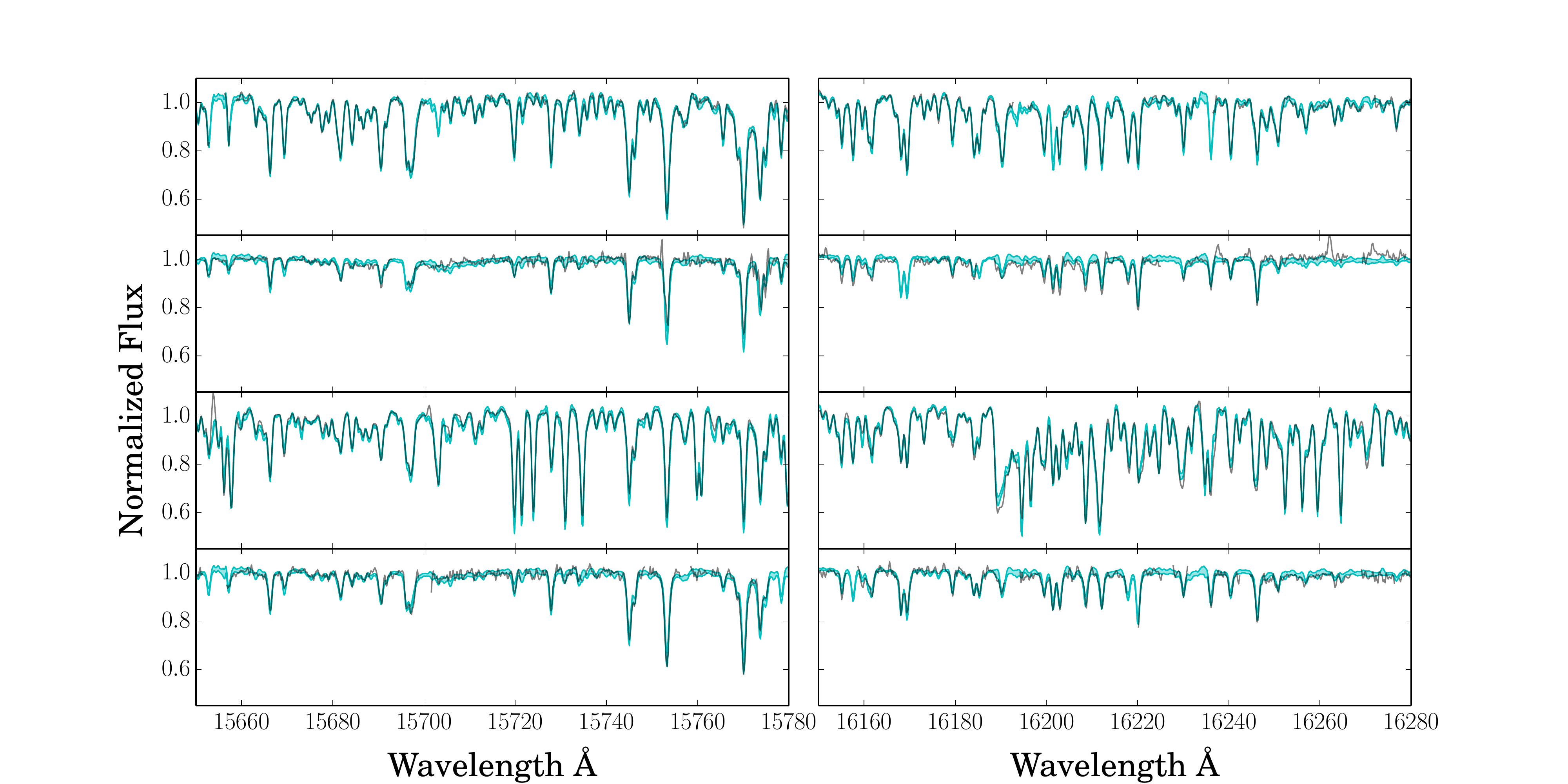}
\caption{The four stars in Figure 3 across narrow wavelength intervals A and B, as described in Section 2.3. The spectra of the stars are plotted in black with the models in cyan, including the span of the scatter of the fit, generated by \tc.}
\label{fig:modeldata}
\end{figure}

\subsection{Identification of \apogee\ Continuum Pixels}
\label{sec:ApogeeContinuum}

The continuum pixels shown in \figurename~\ref{fig:coeffs} for wavelength regions A and B, 
have been determined from the training step with a quadratic-in-labels model operating on spectra normalised by their preliminary pseudo-continuum, using the coefficients returned (see \sectionname~\ref{sec:ContNorm}). About 35\% of the pixels in the resulting baseline spectrum  (the vector $\theta^0_\lambda$) have flux levels within 1\% of unity. However, not all these pixels are suitable continuum pixels, as many of them have significant dependencies,  $\theta^{1,2,3}_\lambda$,  on the three labels. In practice, a good set of continuum pixels can be identified from the \apogee\ spectra using a flux cut in the baseline spectra of the model, 1 $\pm$ 0.15 (0.985-1.015), combined with the smallest 20 - 30 percentile of the first order coefficients, $\theta^{1,2,3}_\lambda$,  which retains between 5-9\% of pixels. We found empirically that changing the latter percentiles to ($\theta^{1}_\lambda$,$\theta^{2}_\lambda$,$\theta^{3}_\lambda$) $<$ (1e$^{-5}$, 0.0045, 0.0085) returns only 6.5\% of the pixels, but ultimately makes for an even  better match to the \aspcap\ label scale; we adopt this procedure. We use the inverse variance weighting of these pixels for the corresponding 2nd order Chebyshev polynomial fit, adding an additional error term that is set to 0 for continuum pixels and a large error value for all other pixels so that the new error term $\sigma^{'}_\lambda$ for each pixel becomes:
$\sigma^{'}_\lambda$ = $\sigma_\lambda$ + $\sigma_{0|LARGE}$.

As we show explicitly in \sectionname~\ref{sec:lowSNR}, we find this to provide a robust continuum-normalization across the stars that are within the parameter range of the training set, across all SNR. 

\subsection{\tc 's Label Transfer for \apogee\ DR10}
\label{sec:APOGEE_DR10_comparison}

Going beyond leave-one-out tests on the set of reference objects, we now apply \tc\ to effect a label transfer to the entire \apogee\ DR10.
We take the spectral model built in the \apogee\ cluster stars in \sectionname~\ref{sec:ModelComplexity}, \sectionname~\ref{sec:take-one-out} and \sectionname~\ref{sec:ApogeeContinuum},
and apply the test step with this model to all DR10 spectra.
Remarkably, we are able to reproduce well the \aspcap\ labels for DR10 spectra. We have run \tc\ through all 47,000 stars in 150 fields in DR10 contained in the available \textit{aspcapstar} files as well as an additional 4800 stars in 20 commissioning fields for which no \aspcap\ parameters were provided in DR10, 
made available in the (non pseudo-continuum-normalized) \textit{apStar} files. 
We also have run \tc\ through the additional commissioning stars available in the \apstar\ files across the fields and in total this
 comprises 55,000 DR10 stars in 170 fields. 

These results of using \tc\ to return labels for all DR10 stars is provided online in \tablename~\ref{tab:online}. 
We provide two columns in this table which indicate the label-space returned for each test object, with respect to the reference objects in the training set. We provide an extrapolation flag, EFLAG, in the table, which indicates if the test star lies outside of the label-space of the reference objects (set to 1 if so). We also determine a distance measurement, $d_{{ref}}$ defined as the distance between the test star (with labels  \teff$_{(test)}$,  \logg$_{(test)}$, \feh$_{(test)}$) and the nearest reference object (with labels \teff$_{(ref)}$,  \logg$_{(ref)}$, \feh$_{(ref)}$), normalised to the maximum distance, so values lie between 0 and 1 (see Equation~\ref{eq:dist}).
Additionally, in the online version of this table, we include a number of  important \aspcap\ flags in the online table. Stars with \badstar\ flag set (in \aspcapflag) may have unphysical stellar parameters and commissioning stars are marked with ``C''. The fidelity of the commissioning stars is uncertain given their different LSF from survey test and training data. The velocity scatter ($\sigma_v$) from the \aspcap\ results as well as the \textit{APOGEE TARGET2} flag (TARG2) are also included.

\begin{equation}
d_{{ref}} =  \substack{\mbox{argmin}\\{\left\{{\mbox{over all n}}\right\}}}
\sum_{1}^{k}
\left( \frac{\Delta \ell_{kn}}{var_k}\right)^2
 \label{eq:dist}
\end{equation}

where $\Delta \ell_{k,n} $ for the three labels is = ($\teff_{(test)} - \teff_{(ref,n)}$), ($\logg_{(test)} - \logg_{(ref,n)}$), ($\feh_{(test)} - \feh_{(ref,n)}$), 
and $var_k$ is the variance of the reference object ensemble distribution in label $l_k$. \\

For the 28,700 stars with parameters (removing all stars flagged as \badstar) provided in DR10, we find we reproduce the \apogee\ labels as follows: 
\teff\ = +12 K $\pm$ 85 K , \logg\ = --0.04 dex $\pm$ 0.18 dex and \feh\ = +0.01 $\pm$ 0.10 dex in \feh. 
The rms errors are comparable to the error estimates for \apogee\ parameters in \citet{Meszaros2013} 
of $\delta$(\teff) $<$ 150 K, $\delta$(\logg) $<$ 0.2 dex and $\delta$(\feh) $<$ 0.1 dex in. 
The typical internal precision on the measured parameters from \tc\ is $\delta$(\teff) $<$ 5.6 K, $\delta$(\logg) $<$ 0.01 dex and $\delta$(\feh) $<$ 0.006 dex.

\begin{table*}[!h]
\small{
\centering
\caption{Partial column excerpt from the online table of stellar labels (\teff, \logg\ and \feh) determined by \tc\ for the 55,000 stars released in 170 fields from \apogee 's data release DR10. } 
\begin{tabular}{| c | c | c |  c | c | c |  c | c | c | c | c | } 
\hline
\small{star ID}  & \teff\ & \logg\ & \feh\ & $\sigma$(\teff) & $\sigma$(\logg) & $\sigma$(\feh) & $\chi^2$ & {$d_{ref}$} & \tiny{EFLAG} \\
\small{(2MASS)} & K & dex &  dex  & K & dex & dex &  & &   \\    
\hline
\tiny{21353892+4229507} & 4160.4 & 1.62  & 0.05  & 3.24 & 0.008  & 0.004  & 2.59 & 0.03 & 0 \\
\tiny{21354474+4250256} & 4824.4 & 4.41  & 0.15  & 8.8  & 0.013  & 0.005  & 0.83 & 0.13 & 0 \\
\tiny{21354775+4233120} & 4704.1  & 2.50  & 0.05  & 9.2  & 0.022  & 0.011  & 0.83   & 0.03 & 0\\
\tiny{21355458+4222326} & 4837.2  & 2.52  & -0.33 &  6.8 &  0.015 &  0.007 &  1.02   &  0.11 & 0\\
\tiny{21360285+4231145} & 4620.0  & 2.09  & -0.43 &  9.6 &  0.023 &  0.011 &  0.94   & 0.15 & 0 \\
\tiny{21360822+4225525} & 4809.6  & 2.75  & -0.03 &  6.9 &  0.016 &  0.008 &  1.15   & 0.017 & 0\\
\hline
\end{tabular}
\label{tab:online} }
\end{table*}  
 
\begin{figure}[!h]
\centering
  \includegraphics[scale=0.23]{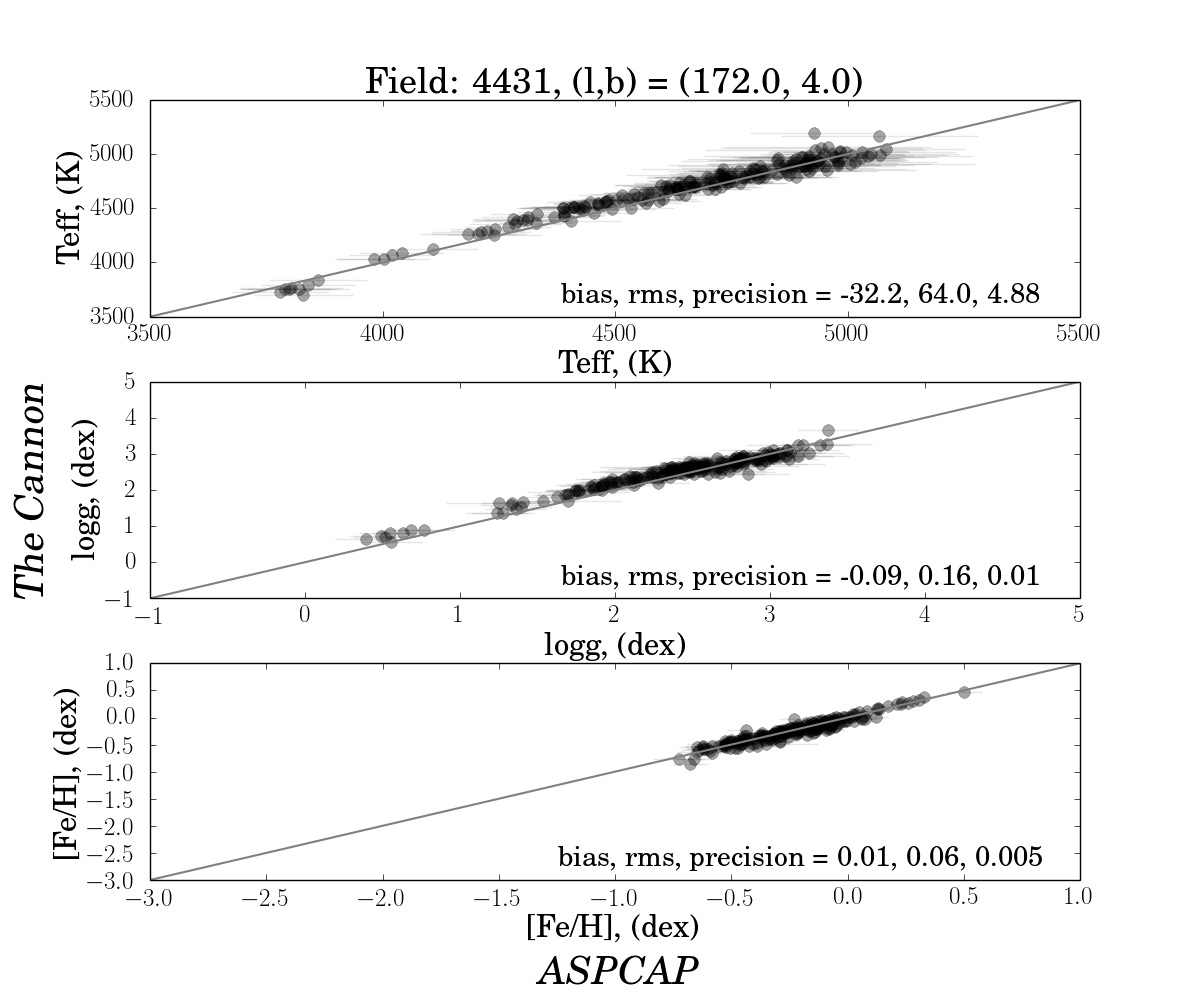}
    \includegraphics[scale=0.23]{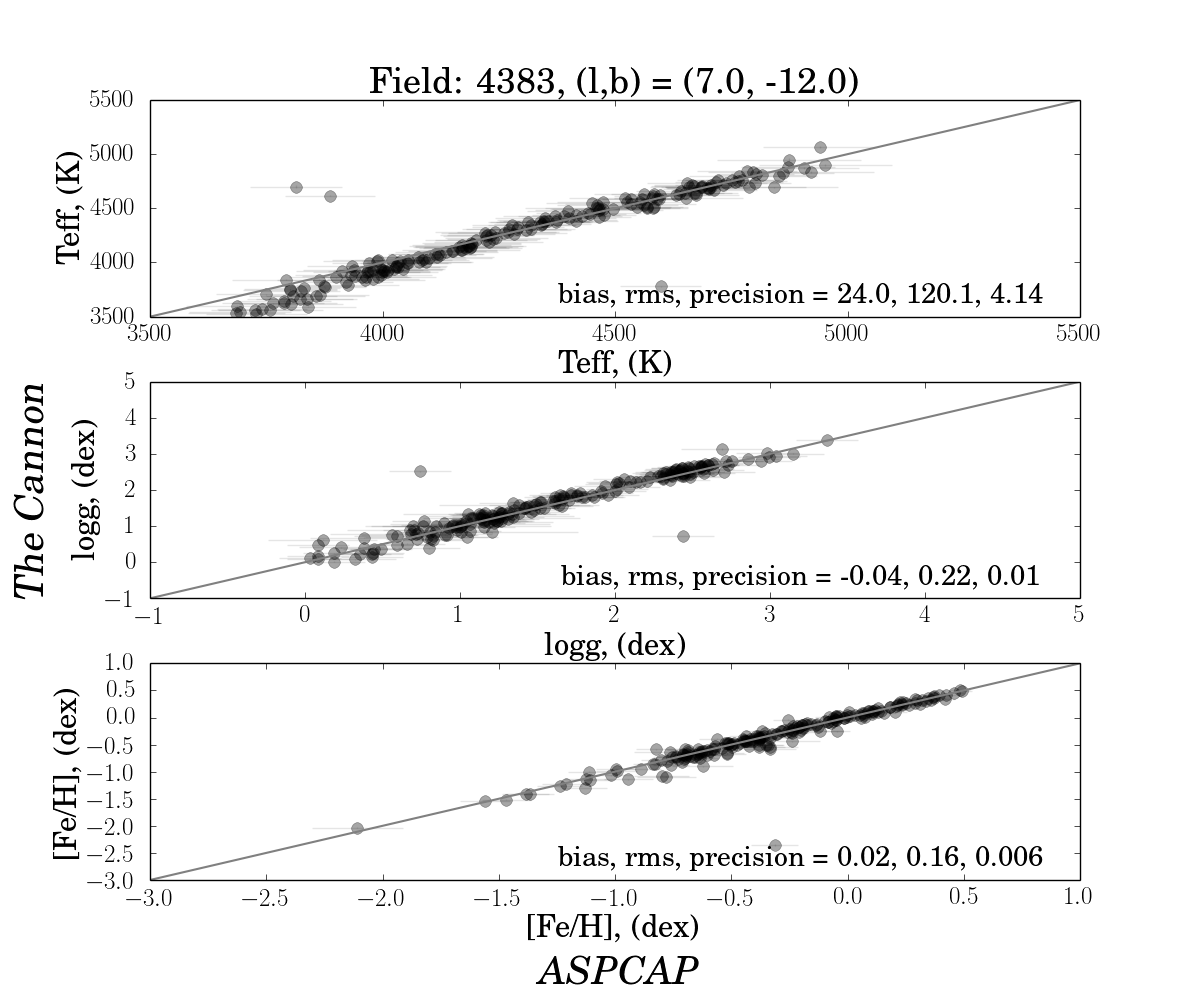} \\
      \includegraphics[scale=0.23]{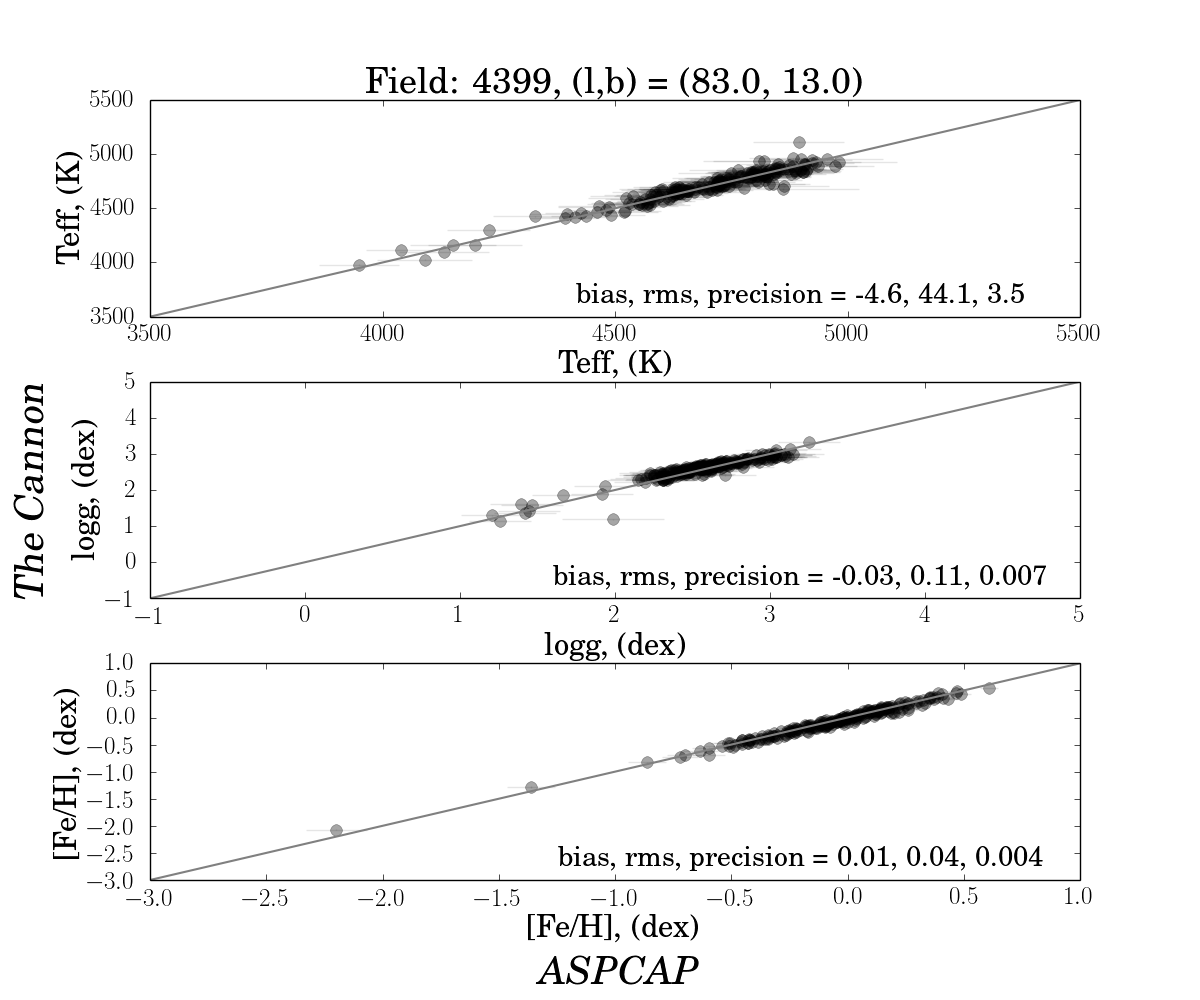}
        \includegraphics[scale=0.23]{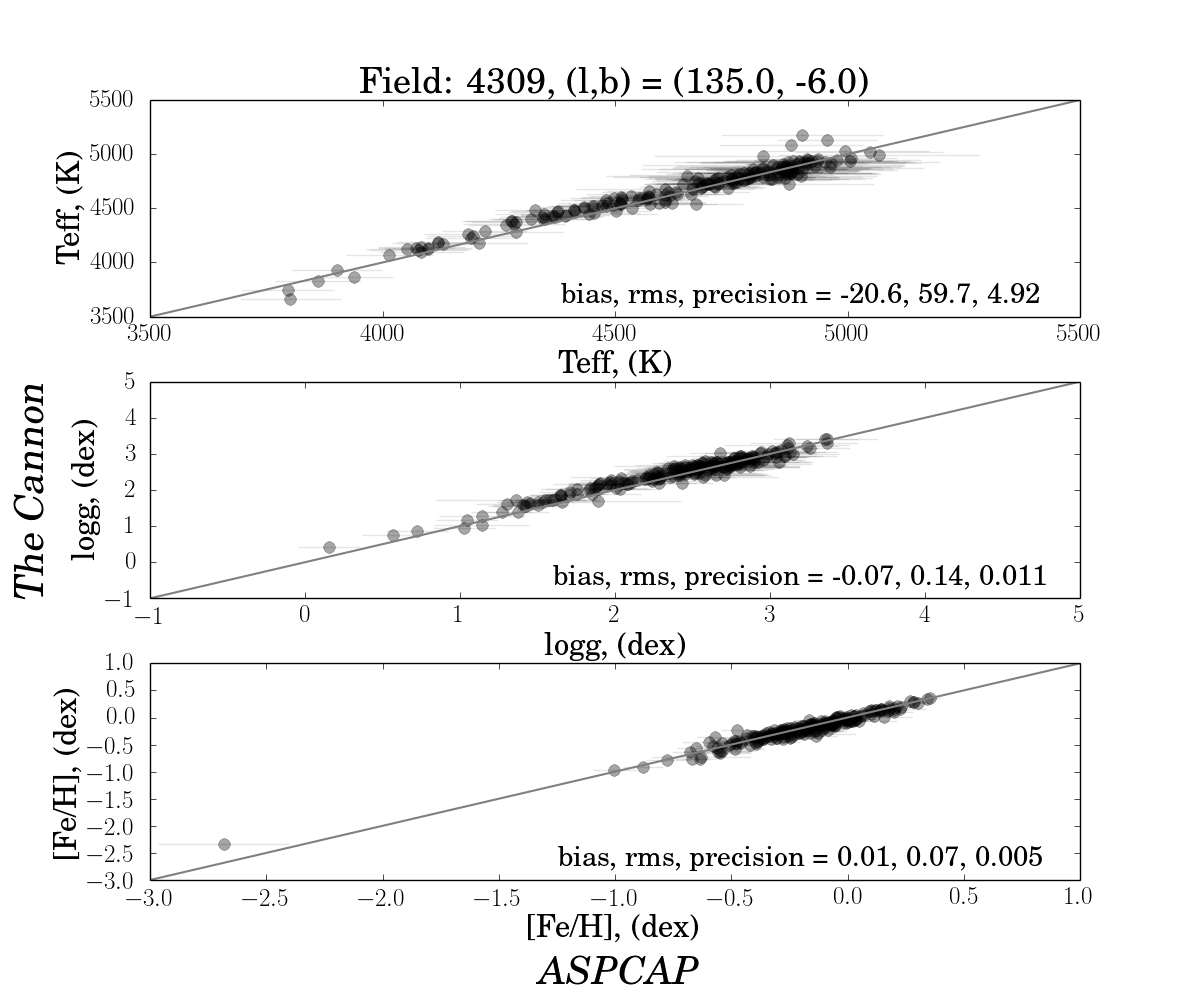} \\
              \includegraphics[scale=0.23]{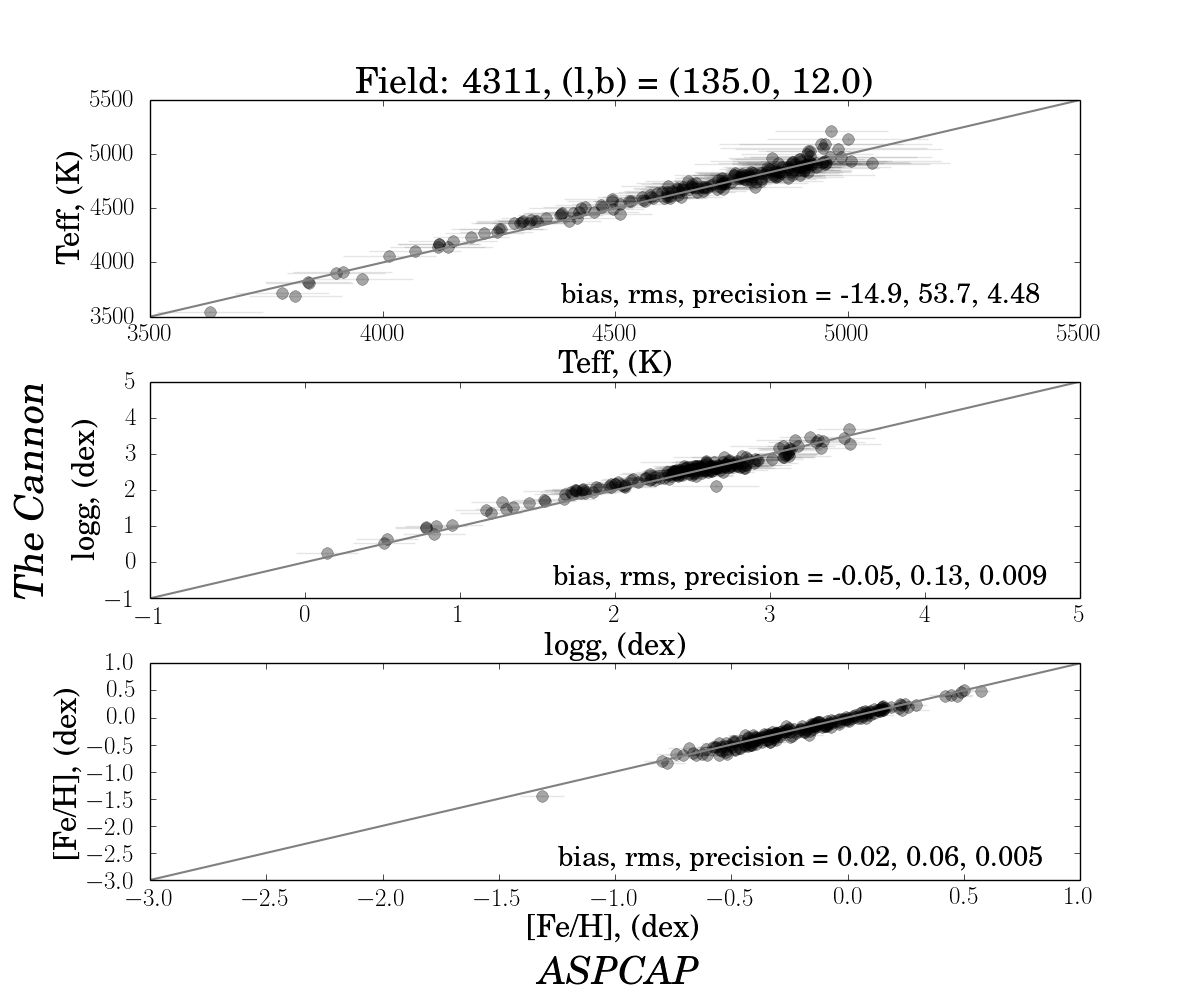}
        \includegraphics[scale=0.23]{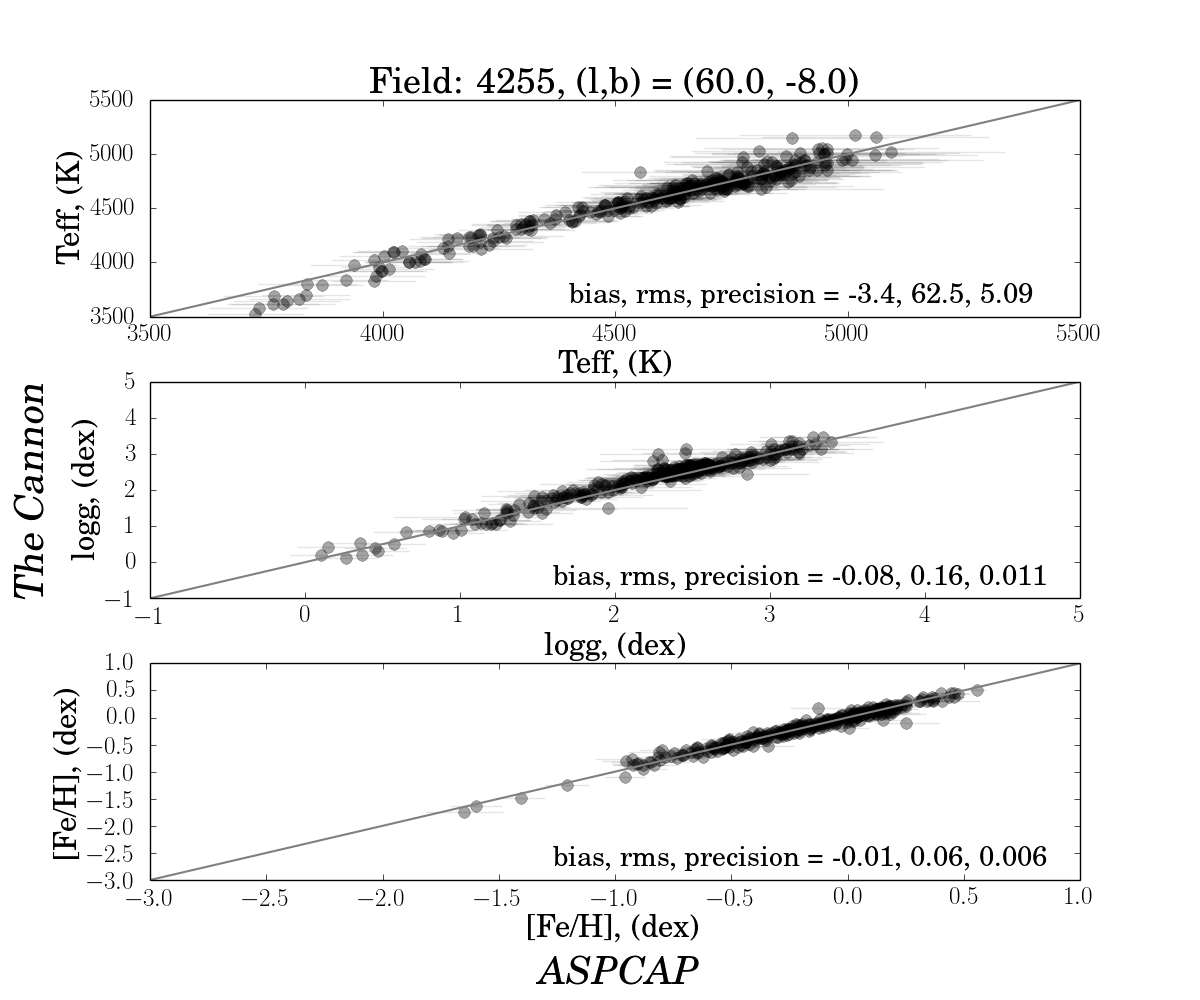} 
\caption{\small{\aspcap\ DR10 versus \tc\ for six different fields including in the disk, bulge and halo. The number of stars is, for each subfigure is 211 (4431), 207 (4384), 217 (4399), 210 (4309), 198 (4311) 319 (4255). Each panel lists the mean difference between the labels (bias), the scatter between the labels (rms) and the formal uncertainly returned by \tc\ (precision). }}
\label{fig:cal}
\end{figure}

The comparison of \tc\ with \aspcap\, showing the bias, rms and formal precision for the labels (\teff , \logg , \feh ) in six sample fields, with bulge, disk and halo targeting, is illustrated in \figurename~\ref{fig:cal}. As for all stars in the survey with \aspcap\ labels, these fields show that we reproduce the \aspcap-corrected stellar parameters with typical rms uncertainties of \teff\ $<$ 100 K, \logg\ $<$ 0.20 dex and $<$ \feh\ $<$ 0.10. These variances are slightly smaller than expected from our cross-validation leave-one-star-out test. This may be because the median of the stellar labels for the DR10 survey object are near the median labels of the reference objects from the training step. 
They are not concentrated to the extreme ends of the range, which have a higher weighting in evaluating the test data with cross validation. 
 \tc 's label transfer also returns values for those $\approx$ 15\% of stars in DR10 are that must be main-sequence dwarf stars. 
These are not shown in \figurename~\ref{fig:cal}, as \apogee\ does not report \aspcap-corrected dwarf parameters for DR10. 
We exclude the stars flagged as bad overall  using the \badstar\ flag from \aspcap\ (this includes stars flagged by \aspcap\ as having bad \teff, bad \logg, high $\chi^2$, an effective temperature more than 1000K from photometric temperature for dereddened color, rotation, SNR low ($<$ 50) or if the parameter is near the \aspcap\ grid edge).  Note we can not return parameters for spectral types not included in our training set (see \sectionname~\ref{sec:AnomalousSpectra}). 

\begin{figure}[!h]
\centering
        \includegraphics[scale=0.35]{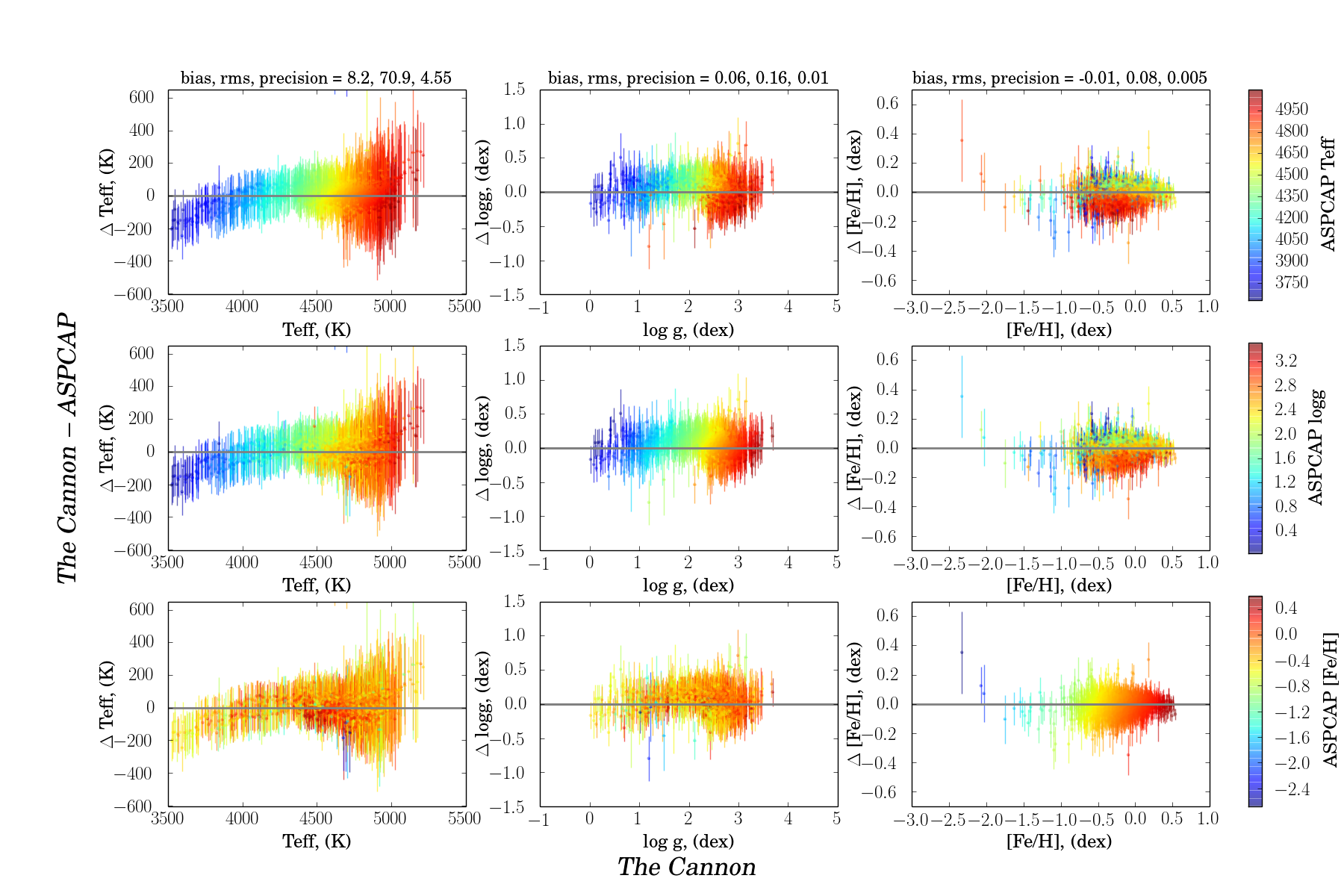} 
\caption{Difference between the labels (\teff, \logg, and \feh) derived by \tc\ and their \aspcap\ DR10 values for all the 1400 stars shown in \figurename~\ref{fig:cal}. The error bars are dominated by those quoted by \aspcap. There are systematic offsets at the coolest temperatures.}
\label{fig:cplot}
\end{figure}

In \figurename~\ref{fig:cal} we show the label differences of \tc\ $-$ \aspcap\ for the 1400 stars from the six sample fields as a function of \aspcap\ \teff, \logg\ and \feh. There are weak trends; at low \teff\ $\sim$ 3700 K, we find temperatures about 100 K cooler than \apogee\ and at low \logg\ we find $\sim$ 0.15 dex larger \logg\ than \apogee. At the lowest metallicities \feh\ $<$ --2.0, we typically report higher metallicities on the order of 0.05 to 0.3 dex. Figure \ref{fig:cplot} emphasizes (compared to
 \figurename~\ref{fig:cal}) the slight systematic deviations at the lowest temperatures. These could be addressed by increasing the complexity of the spectral model; however, they occur beyond the temperature range covered by the reference sets, and hence are inherently less robust (and are flagged as such).

\begin{figure}[!h]
\centering
  \includegraphics[scale=0.26]{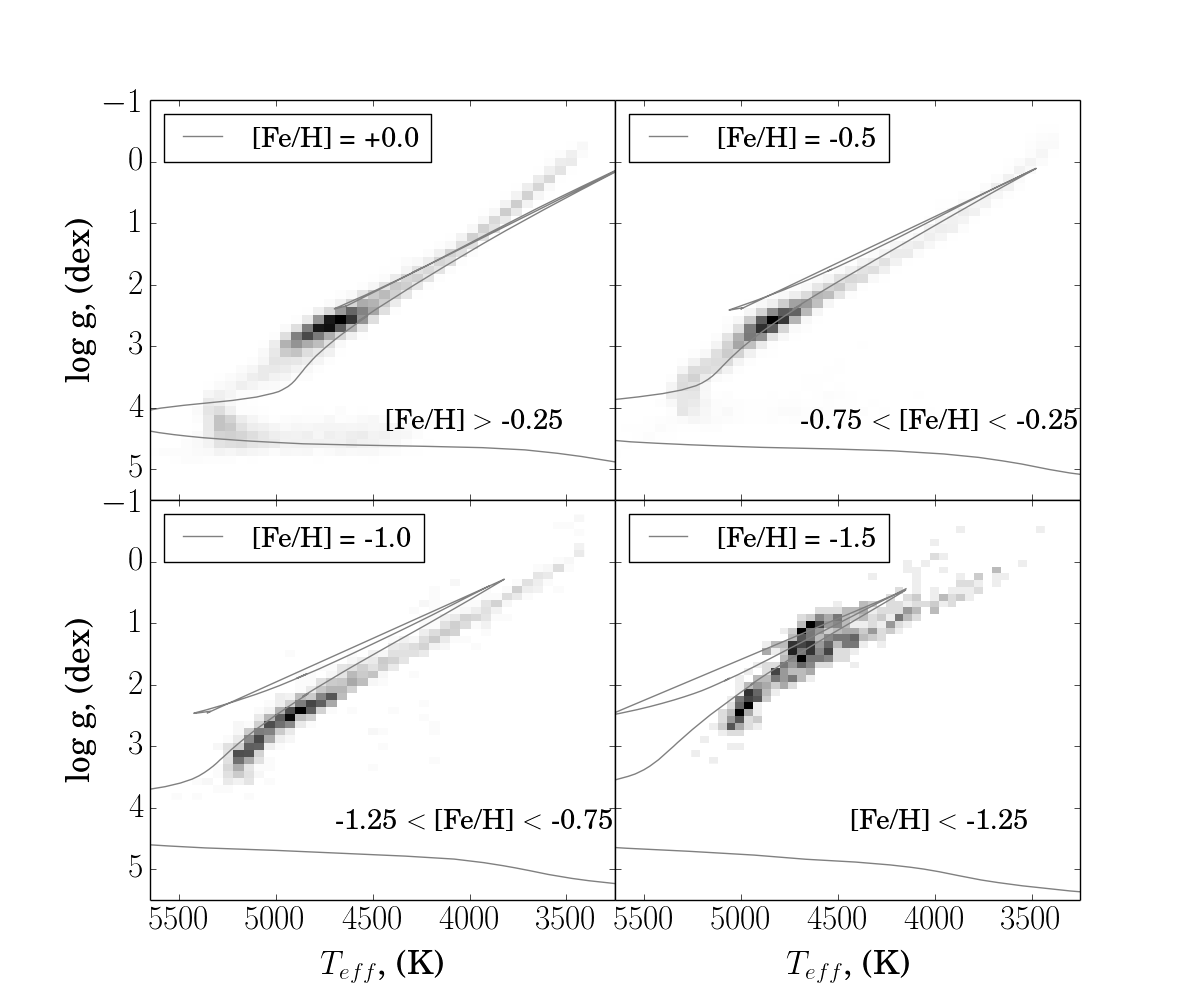}
  \hspace{-20pt}
    \includegraphics[scale=0.26]{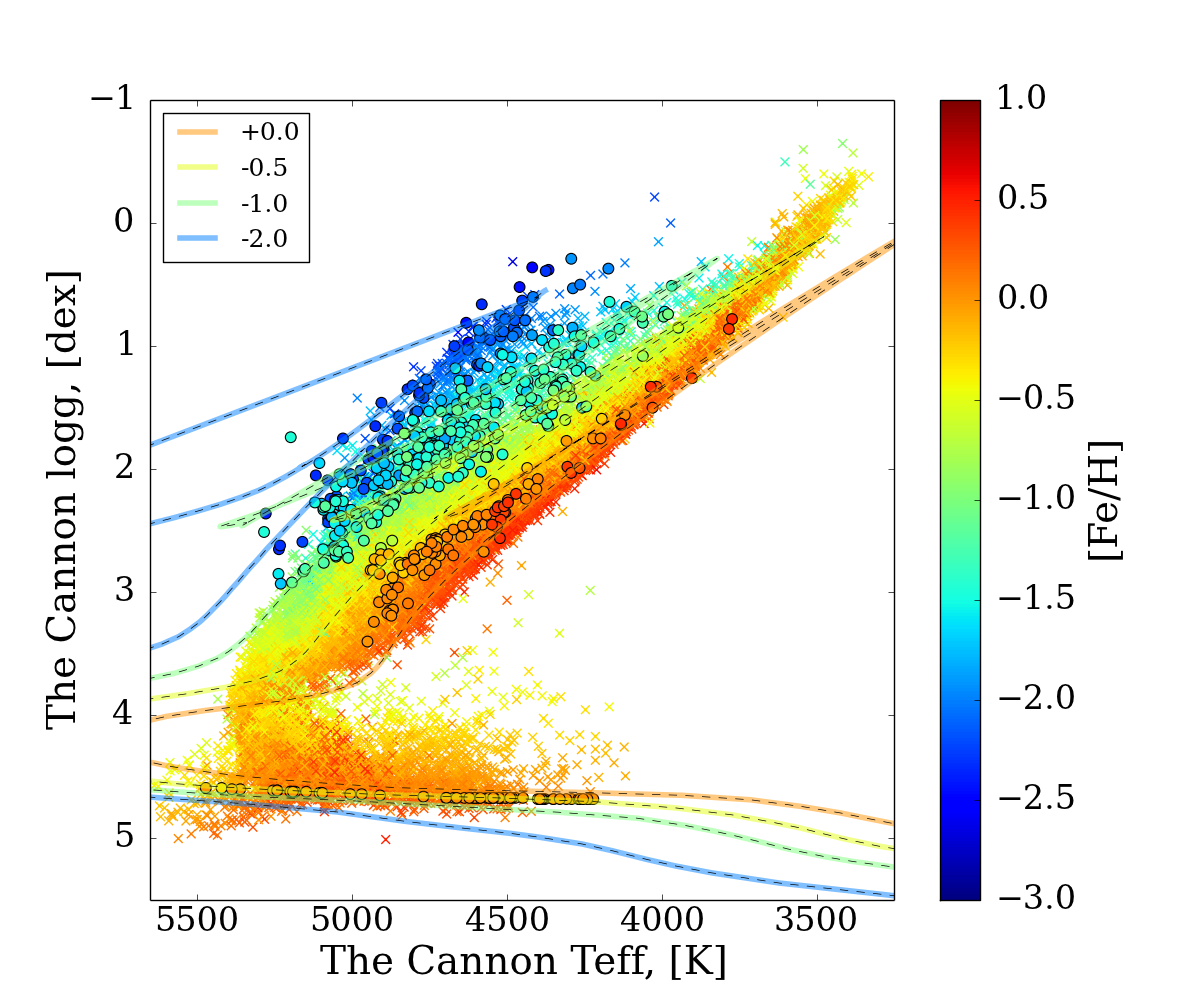}
\caption{Labels for the $\sim$ 35,000 stars from DR10 derived by \tc\ based on \aspcap-corrected labels for the set of reference objects. The set of panels on the left shows \teff-\logg\ in four metallicity bins. There are $\sim$ 19,000, 13,000, 1600, and 1000 stars in the most metal-rich to metal-poor metallicity bins, respectively. The isochrones plotted are 10 Gyr Padova isochrones at the metallicities marked in the upper left hand corners of each sub-panel.  The panel on the right shows all stars coloured in \feh\ on the four isochrones. Note that the \logg\ distribution at low \logg\ is narrow and offset from the giant branch. Reference objects are shown as open circles.} 
\label{fig:iso}
\end{figure}

\begin{figure}[!h]
\centering
 \includegraphics[scale=0.26]{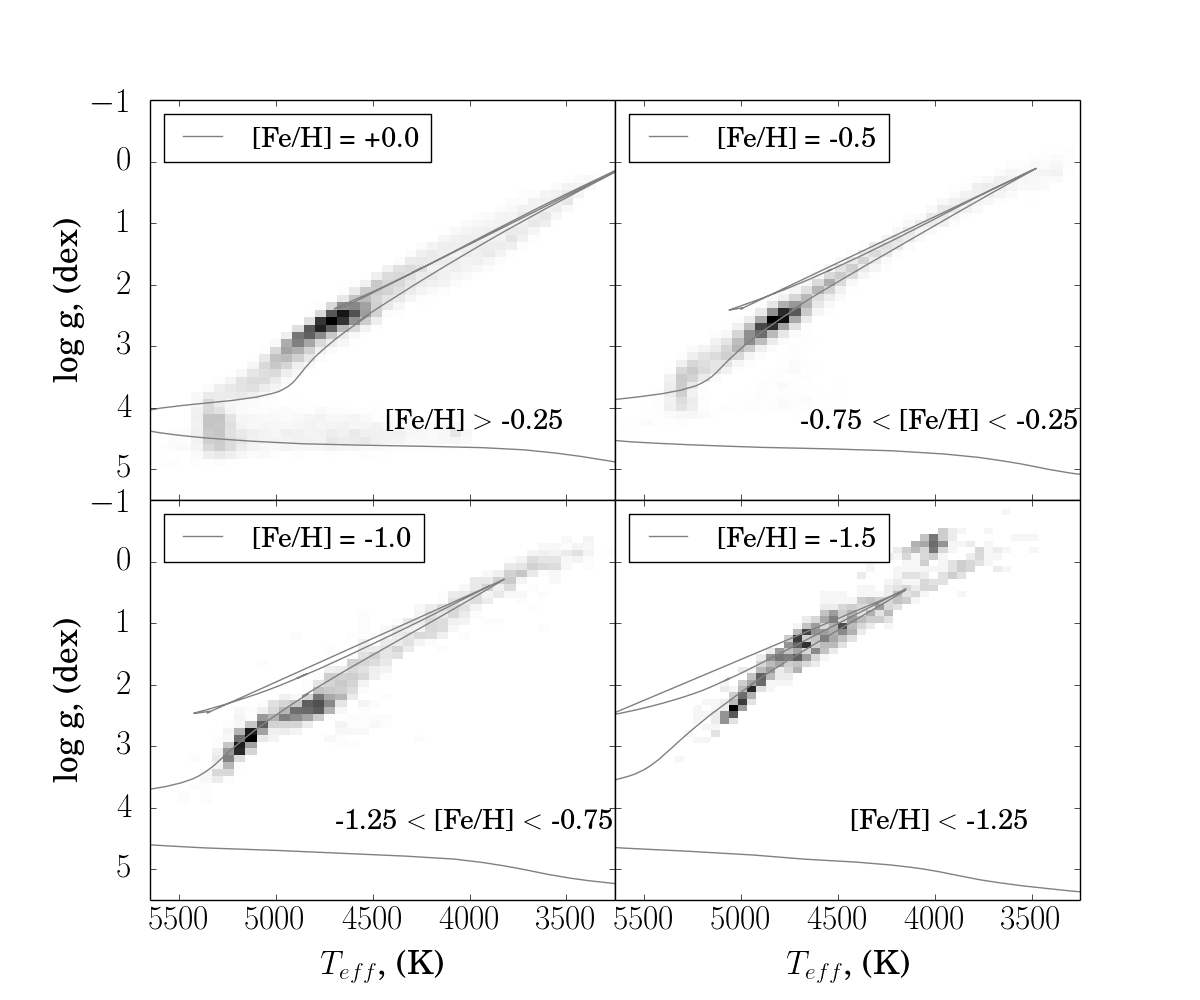}
  \hspace{-20pt}
    \includegraphics[scale=0.26]{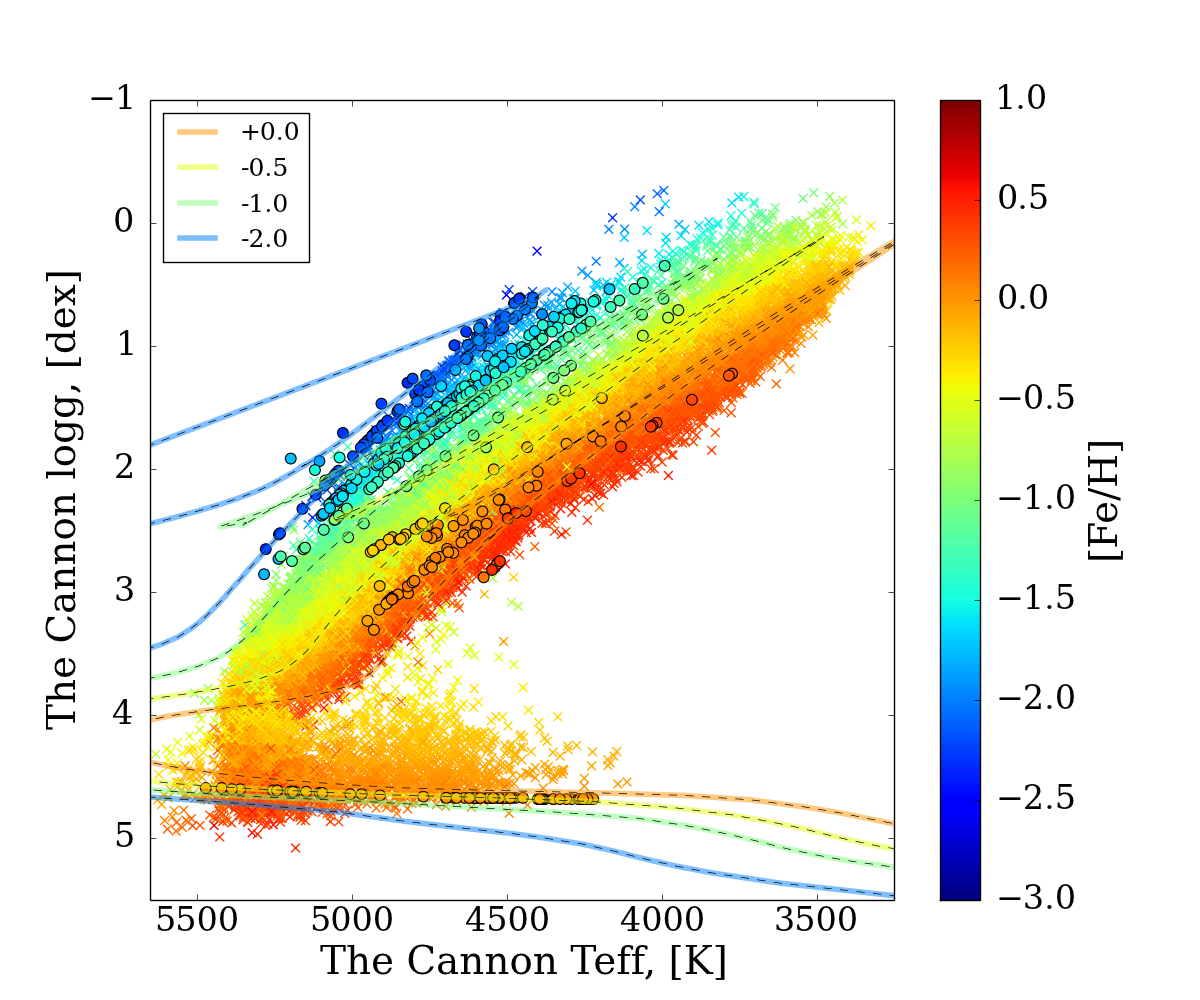}
\caption{Same as \figurename~\ref{fig:iso} but based on the ``isochrone-corrected'' labels for the reference objects. In this case, the labels follow the red giant branch on the isochrones. Note that there is nothing in the mathematics of \tc\ (Equation \ref{eq:specmodel}) that forces resulting labels to lie in physically plausible locations in label space. This is illustrated by the tiny fraction of objects that lie between the main sequence and the giant branch. That most labels lie in physically sensible portions of the \teff-\logg\ plane is a testament to both the quality of the label coverage in the set of reference objects and to the power of \tc\ approach. This is all the more remarkable given that there are basically no main sequence stars among the reference objects.  Reference objects are shown as open circles.}
\label{fig:iso2}
\end{figure}

We show \tc 's resulting label distribution in the \teff-\logg\ plane from \tc\ for the stars in DR10 in \figurename~\ref{fig:iso}. 
This \figurename\ shows the result when \aspcap -corrected labels are used for the reference objects in the training step; \figurename~\ref{fig:iso2} shows the analogous results but for isochrone-corrected reference labels. There are 35,000 stars in these \figurenames\ that remain after excluding stars with the \badstar\ flag set, with velocity scatter $>$ 10 \kms\ and telluric calibration target set. These \figurenames\ also show the labels for the 15\% stars with \logg $>$ 4 dex that must be main sequence stars. 

In short, for all stars with good \aspcap\ labels, we find excellent agreement between \tc\ and \aspcap\ by adopting ASCPAP corrected labels in the training step.
In addition, we are able to derive plausible parameters for dwarf stars in DR10. However, the \teff-\logg\ plane for these stars shows a deviation from the giant branch of the isochrone at low \logg\ (see the right panel of \figurename~\ref{fig:iso}). This is a consequence of the input labels of the training spectra. 

If instead we use the isochrone-corrected \logg\ labels to fix \tc 's spectral model
(\sectionname~\ref{sec:ReferenceObjects}), the results of the label transfer deviate slightly from the \aspcap\ scale in each of the parameters. 
However, with these new \logg\ labels, we find a broad giant branch width that is consistent with expectations in \teff-\logg\ space 
given the metallicity of these stars (see the right hand panel of \figurename~\ref{fig:iso2}). 

This comparison again illustrates both the power of \tc\ to transfer labels, but also its dependence on the choice of suitable reference labels. 

Currently, no priors are incorporated in \tc\ to place the resulting label estimates near physically plausible isochrones. 
Nonetheless, almost all stars lie in physical spaces on the isochrones as shown in \figurenames~\ref{fig:iso} and \ref{fig:iso2} validates the labels. 
The labels for the main sequence stars are presumably much more poorly determined, given the limitations of the reference objects in the training step. 
Remarkably, though, only a handful of stars at low \feh\ and low \logg\ do not lie near conceivable isochrones. 

At metallicities \feh\ $<$ --0.25 the red clump is offset too high in the \logg\ label. This is noted in \citet{bovy2014} who estimate that this offset shifts the red clump and red giant branches 0.2 dex closer together. 
Our \logg\ labels in \figurename~\ref{fig:iso} are essentially identical to \apogee\ \aspcap\ labels (offset -0.04 dex in \logg\ for DR10) and the left panels show that the red clump stars which should be seen as a density maxima of stars around \logg\ $\sim$ 2.5, \teff\ $\sim$ 
are offset to higher \logg\ than the red clump branch of the Padova isochrone, for stars \feh\ $<$ --0.25. 
This offset is on the order of 0.2 dex at \feh\ = -0.5 and is present for both \aspcap-corrected labels and isochrone-corrected labels. 
This may indicate that the \aspcap\ temperature scale is offset too cool in DR10 (but as a function of \feh).

\begin{figure}[!h]
\centering
  \includegraphics[scale=0.26]{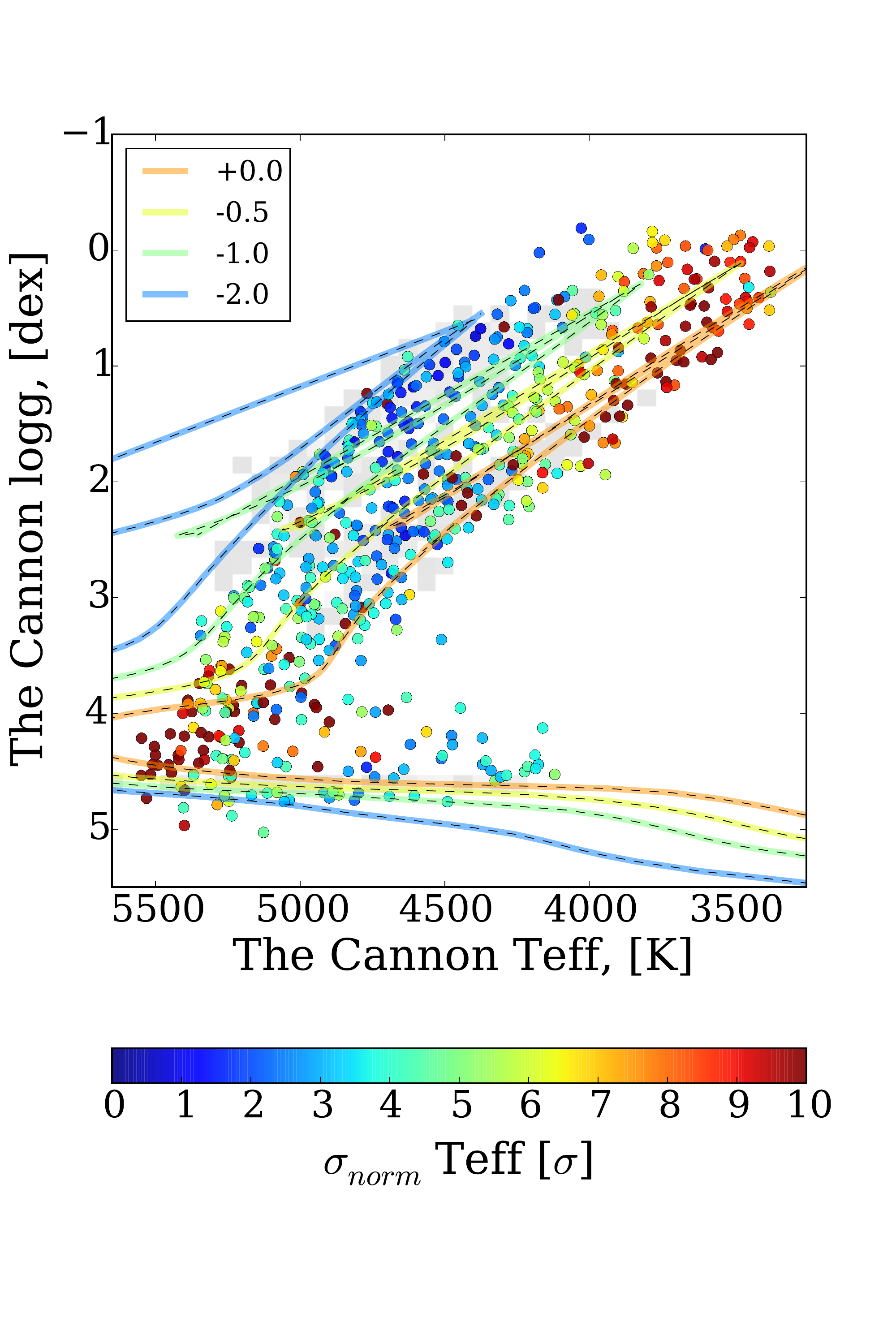}
    \includegraphics[scale=0.26]{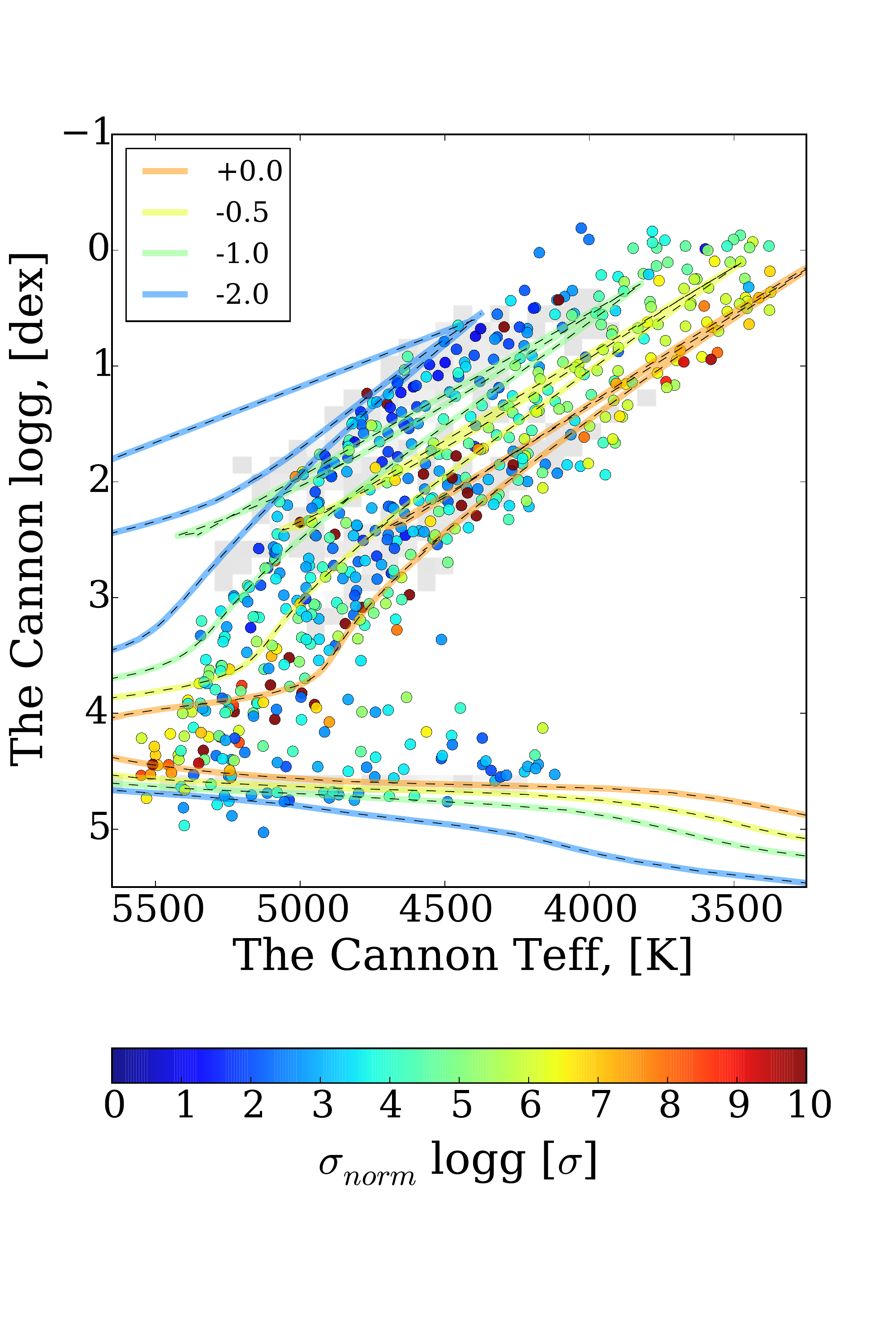}
    \includegraphics[scale=0.26]{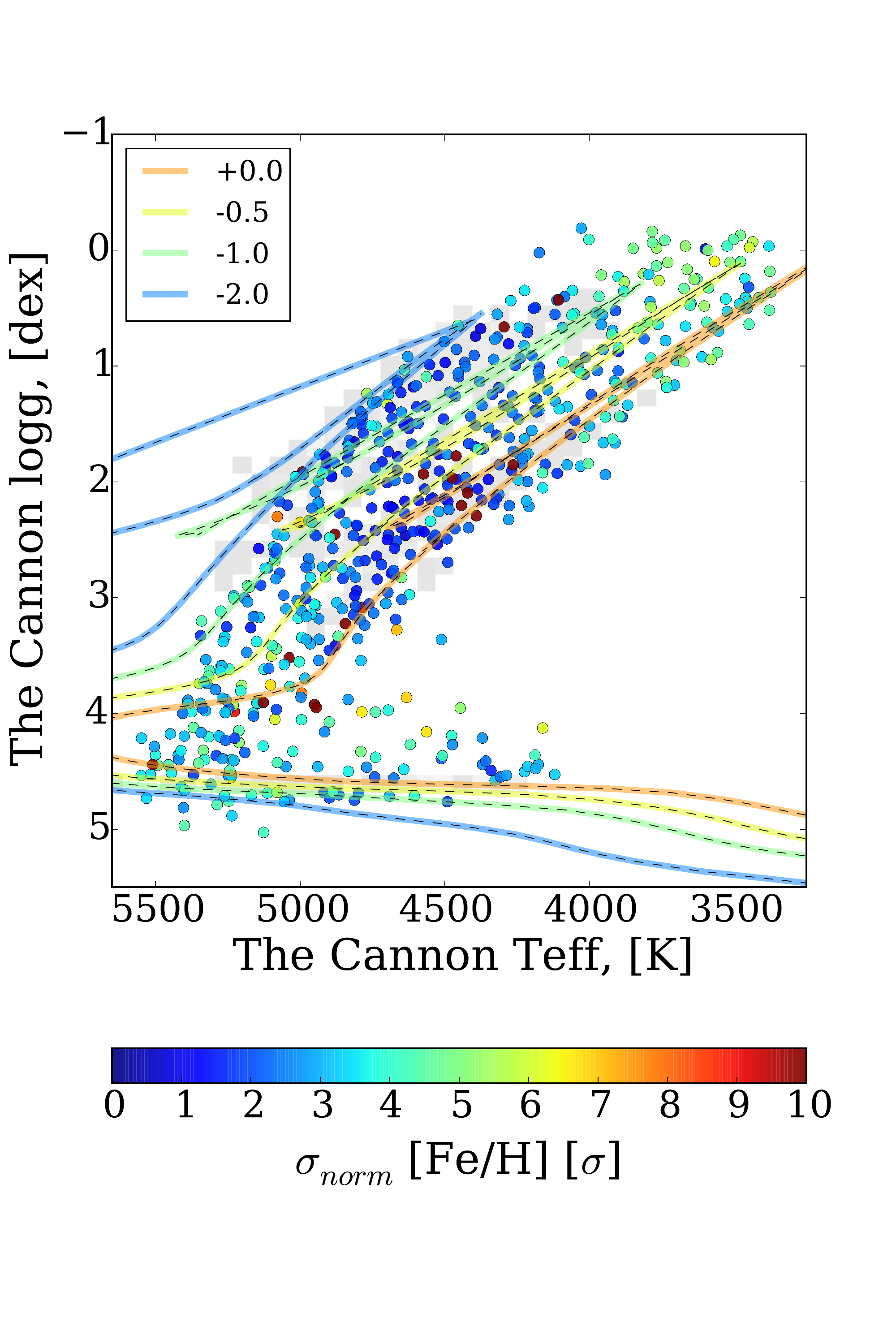}
    \caption{The standard deviation in the labels returned in \teff, \logg\ and \feh, shown in the \teff$-$\logg\ plane, normalised by the optimisation error on each measurement, for 20 bootstrapping tests of the training set. The representative sample of $\sim$ 670 stars shown here has been drawn from an equal sampling of a grid spaced by 100 K in \teff, 0.25 dex in \logg\ and 0.25 dex in \feh\ from the labels returned using the model trained on the isochrone-corrected reference objects. The location of the reference objects is shown in the grey shaded regions in the panel. Note the narrow region of reference objects also on the main sequence. The highest scatter in the labels is seen for regions where the labels are extrapolated. These figures are shown for the isochrone-corrected labels discussed in Section 2.4.}
    \label{fig:sigma}.
\end{figure}

So far, we have not yet explored the (possibly systematic) uncertainties of the label estimates, arising from the incomplete, and in part sparse coverage of label-space by the reference objects. Though the labels in Figure~\ref{fig:iso2} lie mostly in physically plausible locations, the fidelity of the labels in particular in the extrapolated part of label space must be scrutinized. To do this, we created twenty different spectral models, by bootstrap-sampling from the set of reference objects and ran the training step on each of them. Using these twenty different spectral models, we derived twenty different label estimates for a sub-set of the survey objects;  we picked one survey object in each cell of a three dimensional grid in \teff, \logg\ and \feh, of 100 K, 0.25 dex and 0.25 dex respectively. We then took the dispersion among these label estimates as a diagnostic of the uncertainties arising from the sparseness of the training set; this is shown in Figure \ref{fig:sigma}, normalized by the formal error on each label. The location of the reference objects is indicated in the grey shaded regions. These Figures show that in the parts of label space well-covered by reference objects, these two uncertainties are comparable. In the extrapolated regions of label space (here, the main sequence turn-off and the metal-rich tip of the RGB), the dispersion among the bootstrap-returned labels is considerably higher than the formal uncertainties; here the incomplete coverage of label space by reference objects becomes the dominant uncertainty. 

\begin{figure}[!h]
\centering
  \includegraphics[scale=0.26]{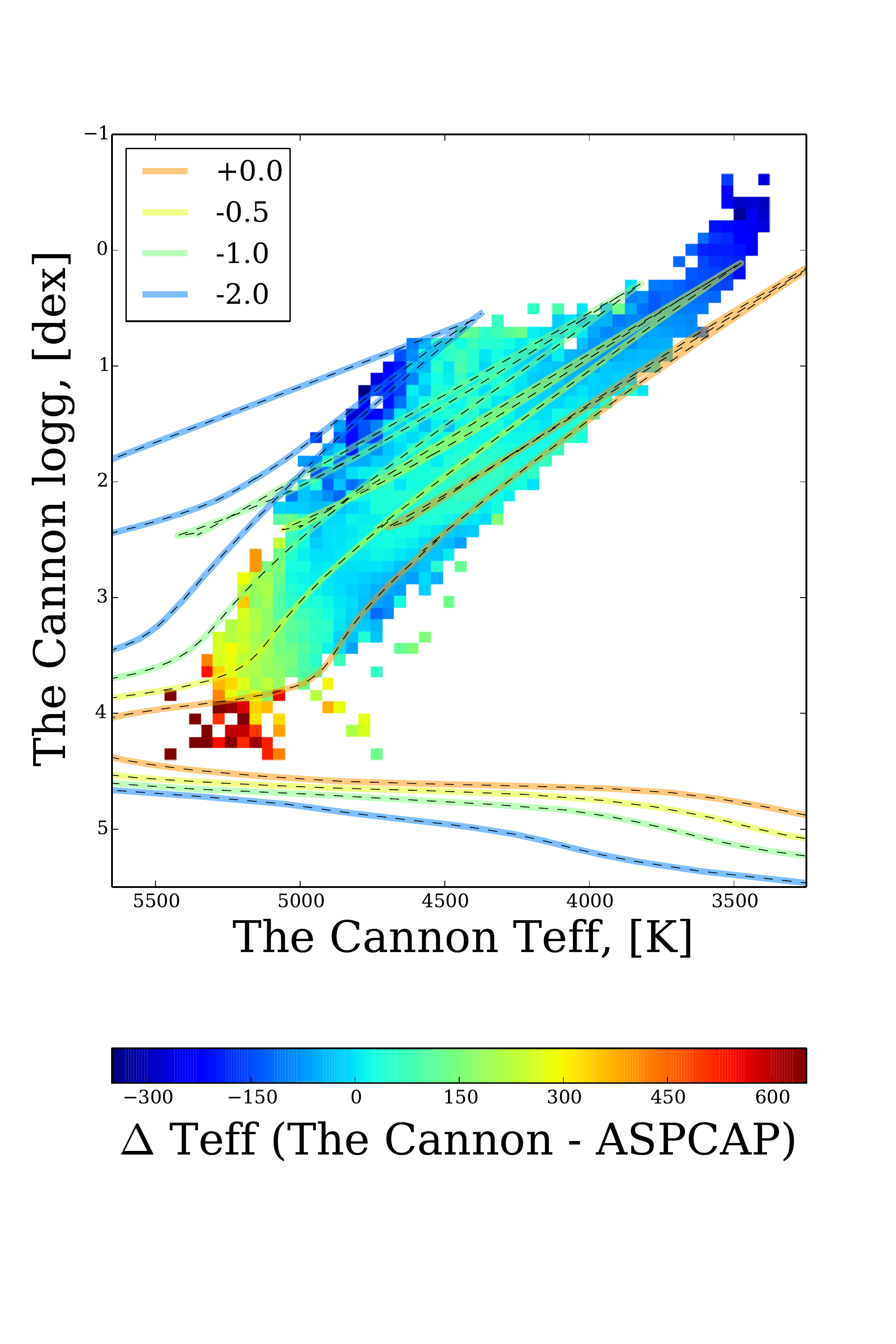}
    \includegraphics[scale=0.26]{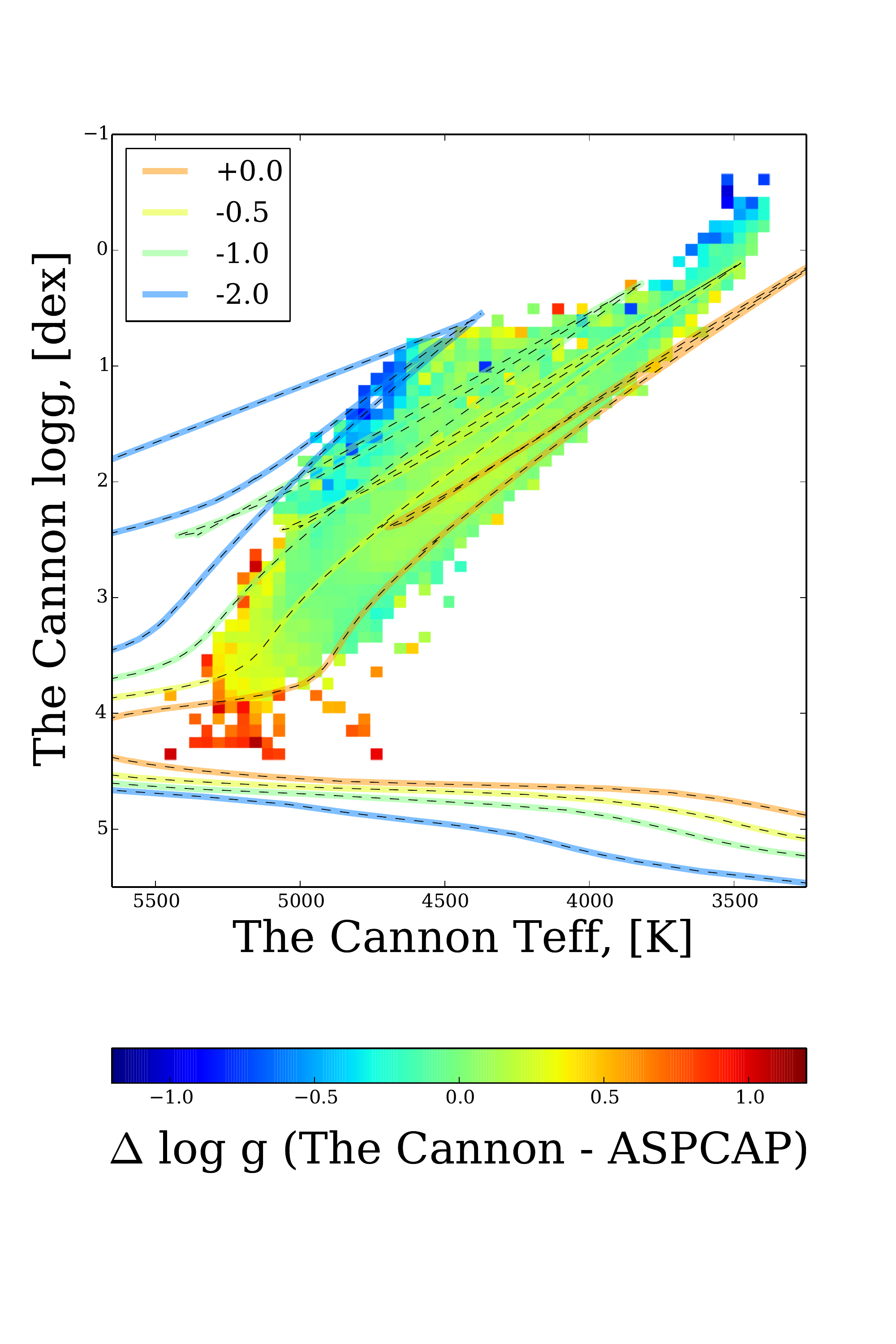}
    \includegraphics[scale=0.26]{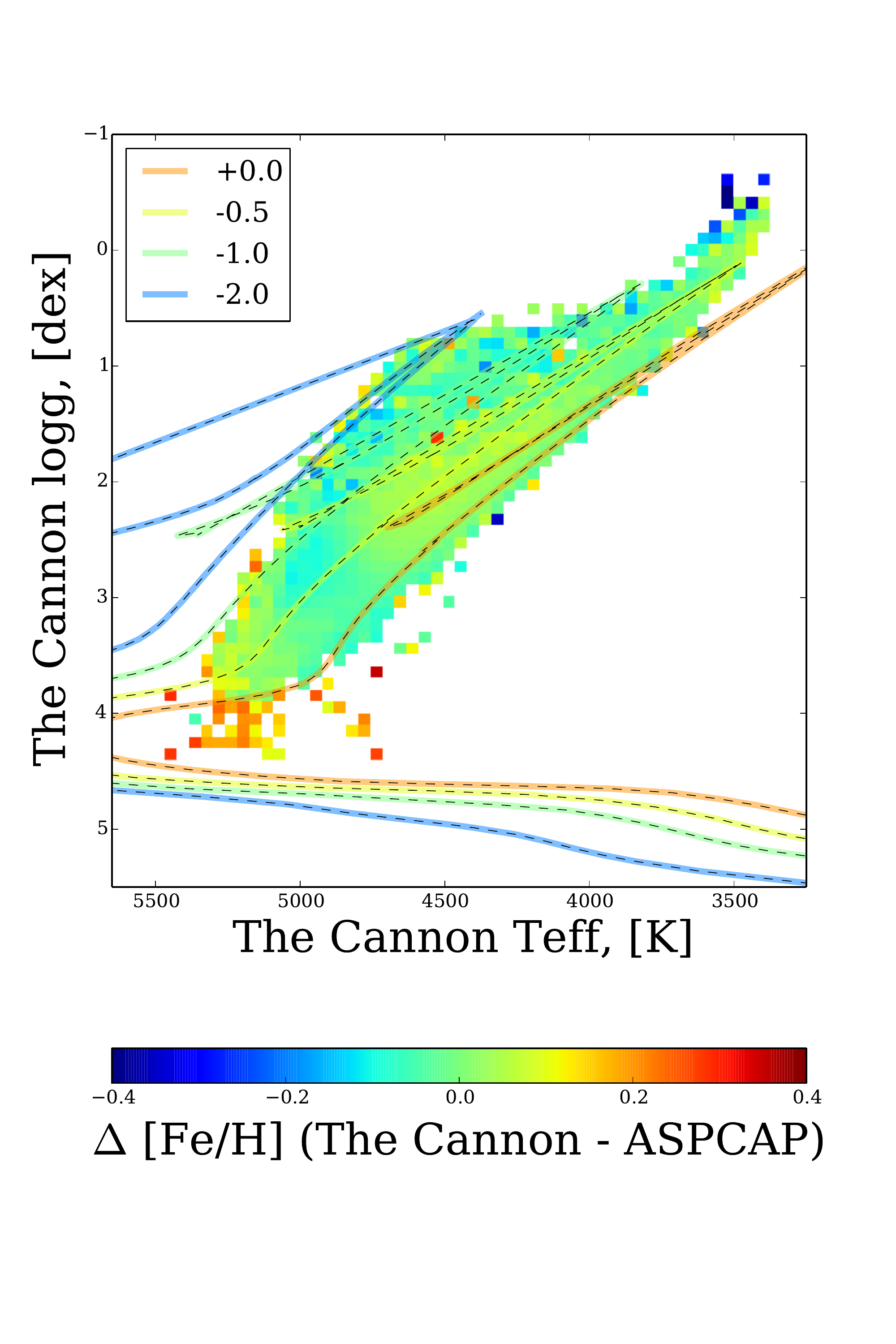}
    \caption{The difference in labels between \tc\ and \aspcap\, indicating the regions of extrapolation where the difference in the labels extends beyond the estimated errors of the methods, due to the limited sampling of the reference objects which does not fully cover the label-space of the survey. The \aspcap-corrected training labels were used to generate the model applied at the test step on the DR10 data.}
    \label{fig:sigma2}.
\end{figure}

Both methods rely on extrapolation of labels, \tc\ uses the extrapolation of the spectral model,  \aspcap\ performs the extrapolation at the label-inference stage (see Meszaros et al., 2013). Figure \ref{fig:sigma2} compares the stellar labels from \tc\ to those from \aspcap. It is only in these extrapolated regions that the labels from \tc\ and \aspcap\ deviate beyond the estimated errors of the \aspcap\ pipeline. Again, in these regions, neither survey is calibrated to empirical ground truth. These figures also show that the deviations between \tc\ and \aspcap\ are systematic and not random. In general, these regions of extrapolation will be directly dependent on the survey and the sets of reference objects.  Figure  \ref{fig:sigma2} highlights the regions of missing label space where stronger constraints on the labels are needed; that is reference objects. In general however, this approach allows the propagation of the uncertainties from imperfect sets of reference objects to the resulting label estimates of the test objects.

\subsection{Failures: Types of Spectra not Represented among the Reference Objects}
\label{sec:AnomalousSpectra}

\begin{figure}[!h]
\centering
\includegraphics[width=\hsize]{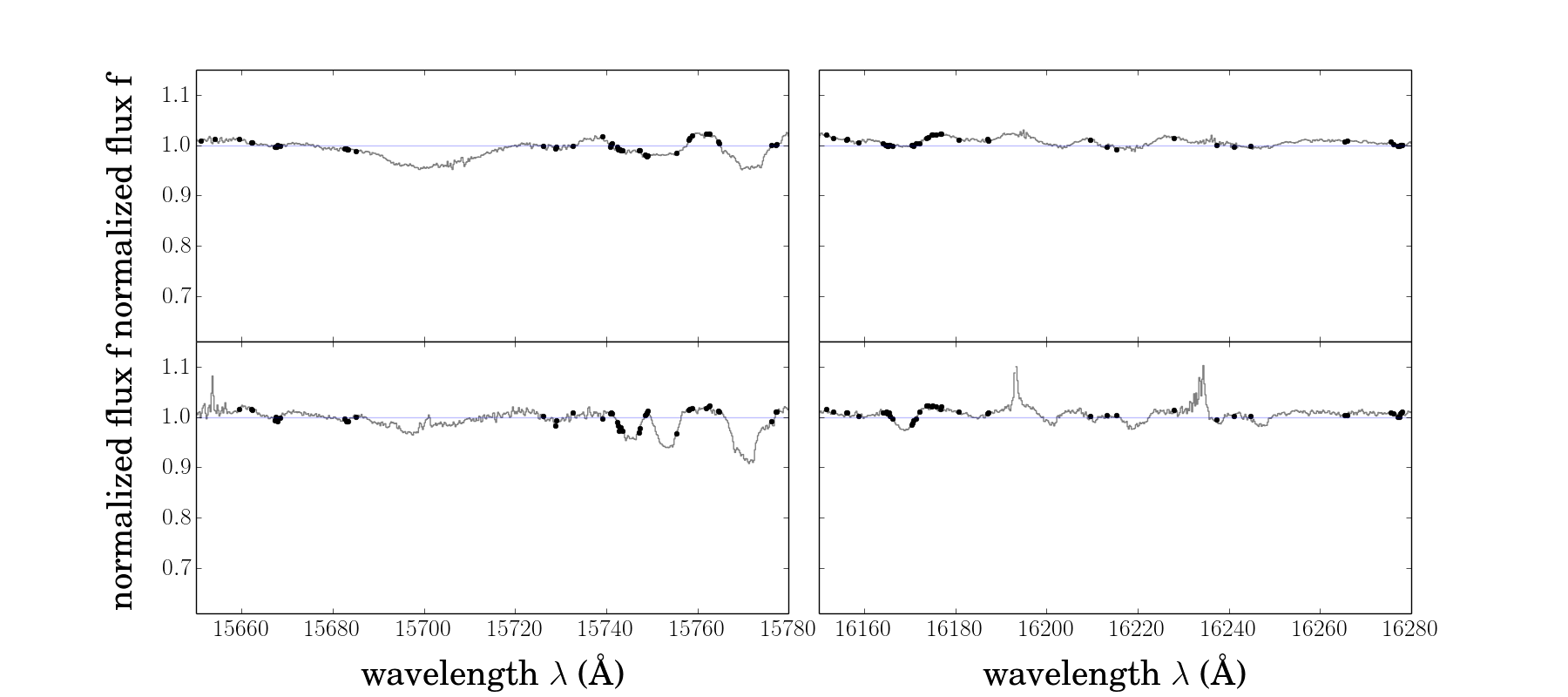}
\caption{Examples of hot rotating dwarfs in the \apogee\ DR10 data across regions A and B, comparable to \figurename~\ref{fig:coeffs}. These types of stars are \textit{not} included in our set of reference objects. Therefore, the label transfer by \tc\ leads to grossly unphysical label estimates.}
\label{fig:dwarfs}
\end{figure}

The dwarf spectra in our reference set only come from the Pleiades cluster, at a single metallicity. 
This restricted sample limits our ability to determine the stellar parameters for dwarf stars. 
Given these training data, our model \textit{can} differentiate dwarfs from giants, as long as their spectra are comparable to that of the Pleiades. 
However, none of the dwarfs in our training set are hot rotating objects with broad line features. 
Three examples of stars with broad line features that are in the test data but not included in our training dataset are shown in \figurename~\ref{fig:dwarfs}.

It is possible to differentiate these stars with \tc\ because they are output in non-physical space in \teff-\logg, and present as a group of very metal poor, \feh\ $\sim$ --2.0, low \logg\ stars $\sim$ 0, with cool temperatures $\sim$ 4000 K. The metal poor solution determined by \tc\ reflects the dearth of lines in the spectra for these hot stars, given the training model. This group of stars is flagged in \aspcap\ with a \rotwarn\ flag set. We therefore are able to exclude these stars from our analysis using this condition.

 \subsection{Performance at modest SNR}
 \label{sec:lowSNR}

\begin{figure}[!h]
  \includegraphics[width=\hsize]{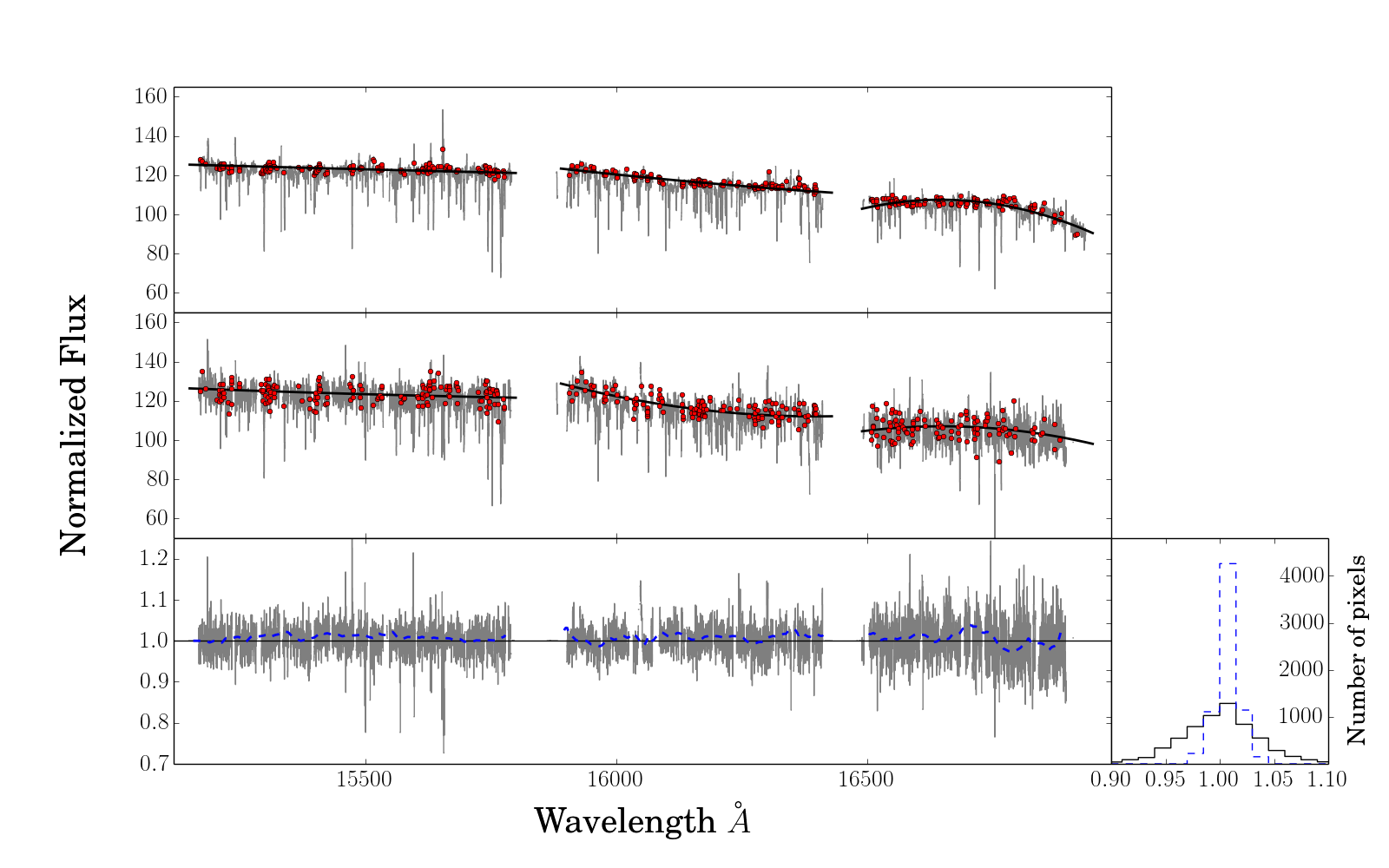}
  \caption{Comparison of the continuum normalization of the same star at high and modest SNR. The \apogee\ \apstar\ combined visit spectra is shown in the top panel (SNR = 120) and the \apstar\ spectra for the 4th visit (SNR = 25) is shown in the second panel. The bottom panel is the ratio of the continuum-normalized spectra of the high and medium SNR spectra and the blue dashed line is a running median of this ratio over 20 \AA, showing a small bias. The histogram of this ratios and of its median are given in at the right of the bottom panel.}
\label{fig:lowsnr}
\end{figure}

\begin{figure}[!h]
\centering
\includegraphics[scale=0.25]{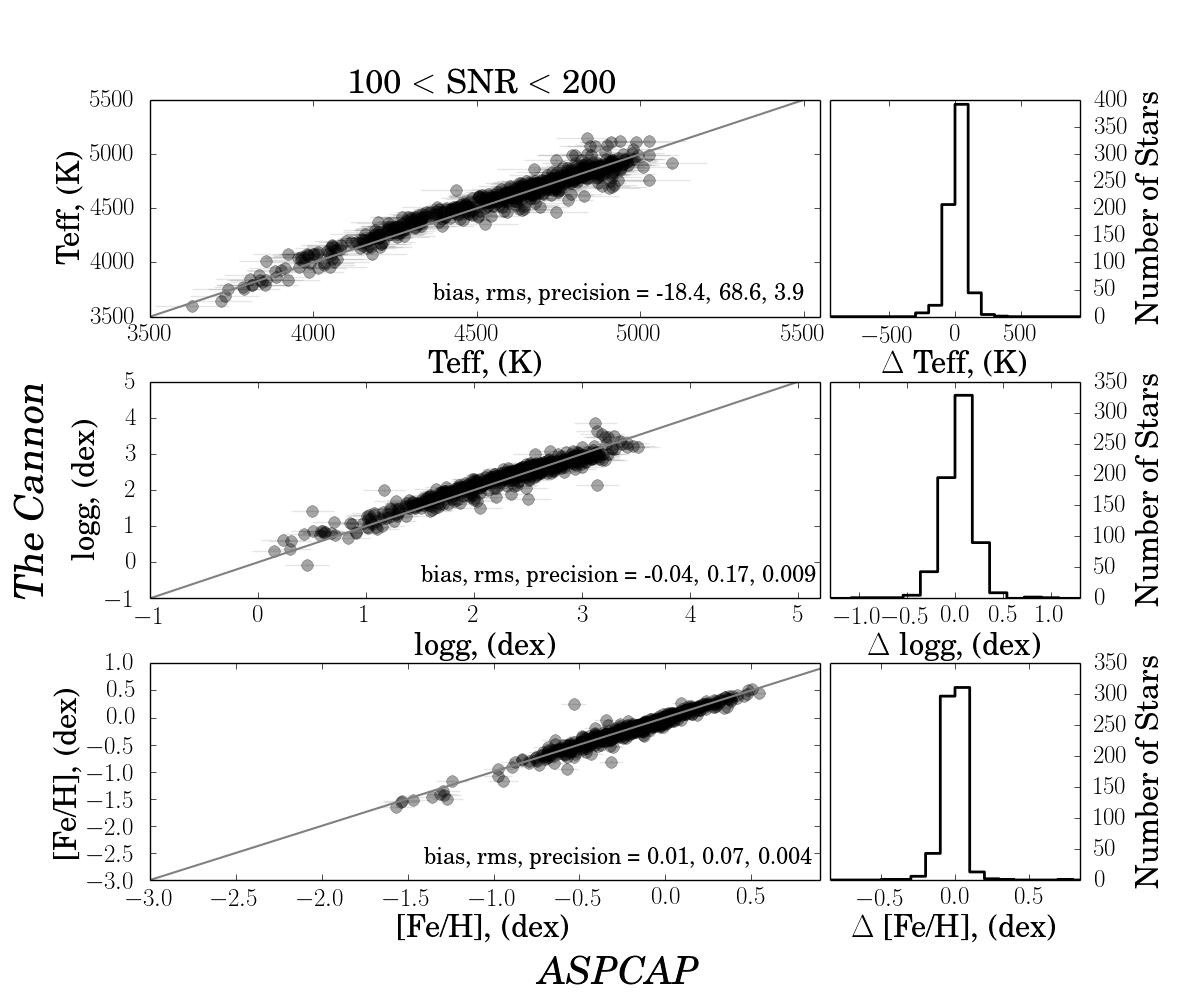}
\includegraphics[scale=0.25]{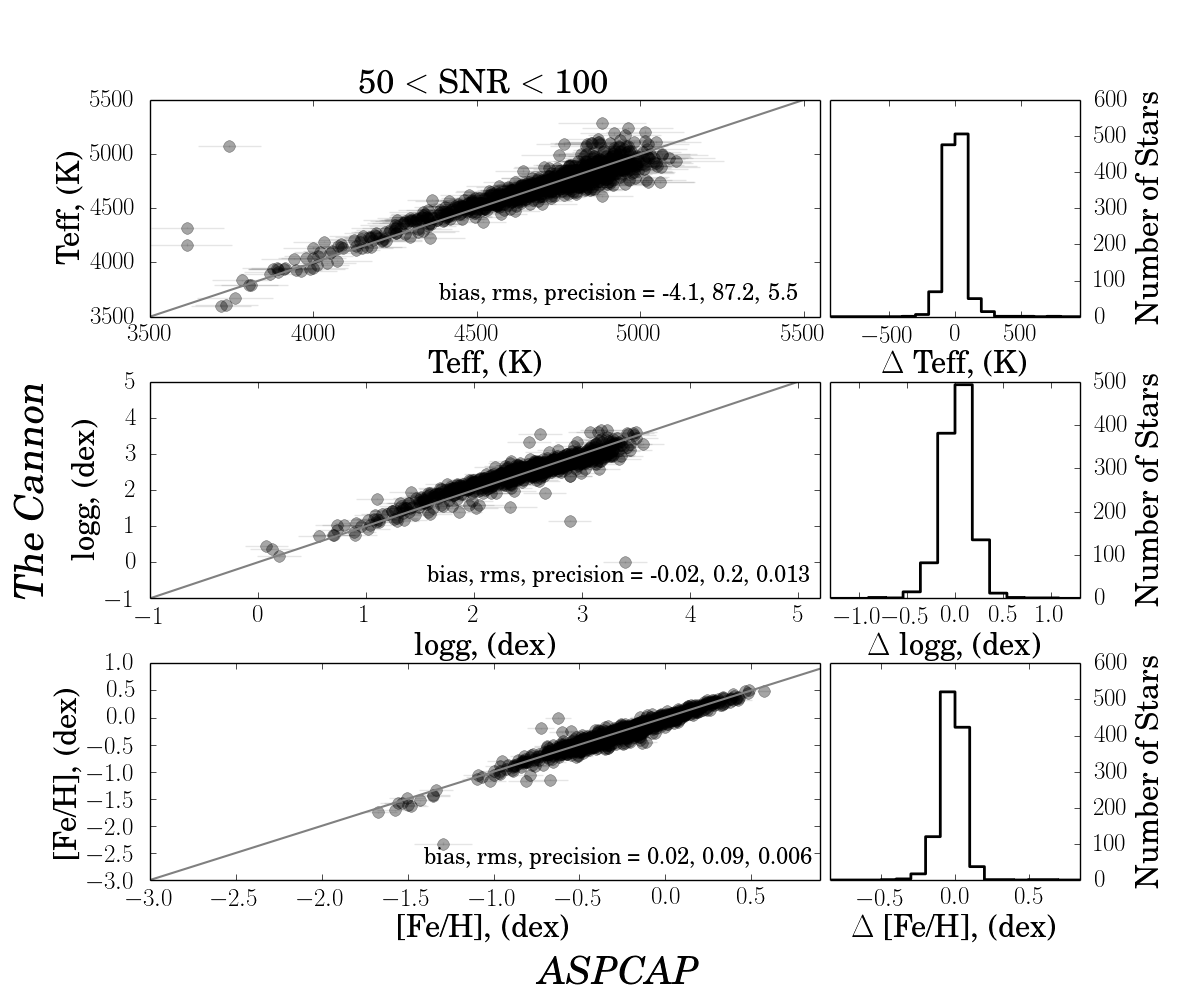}\\
\includegraphics[scale=0.25]{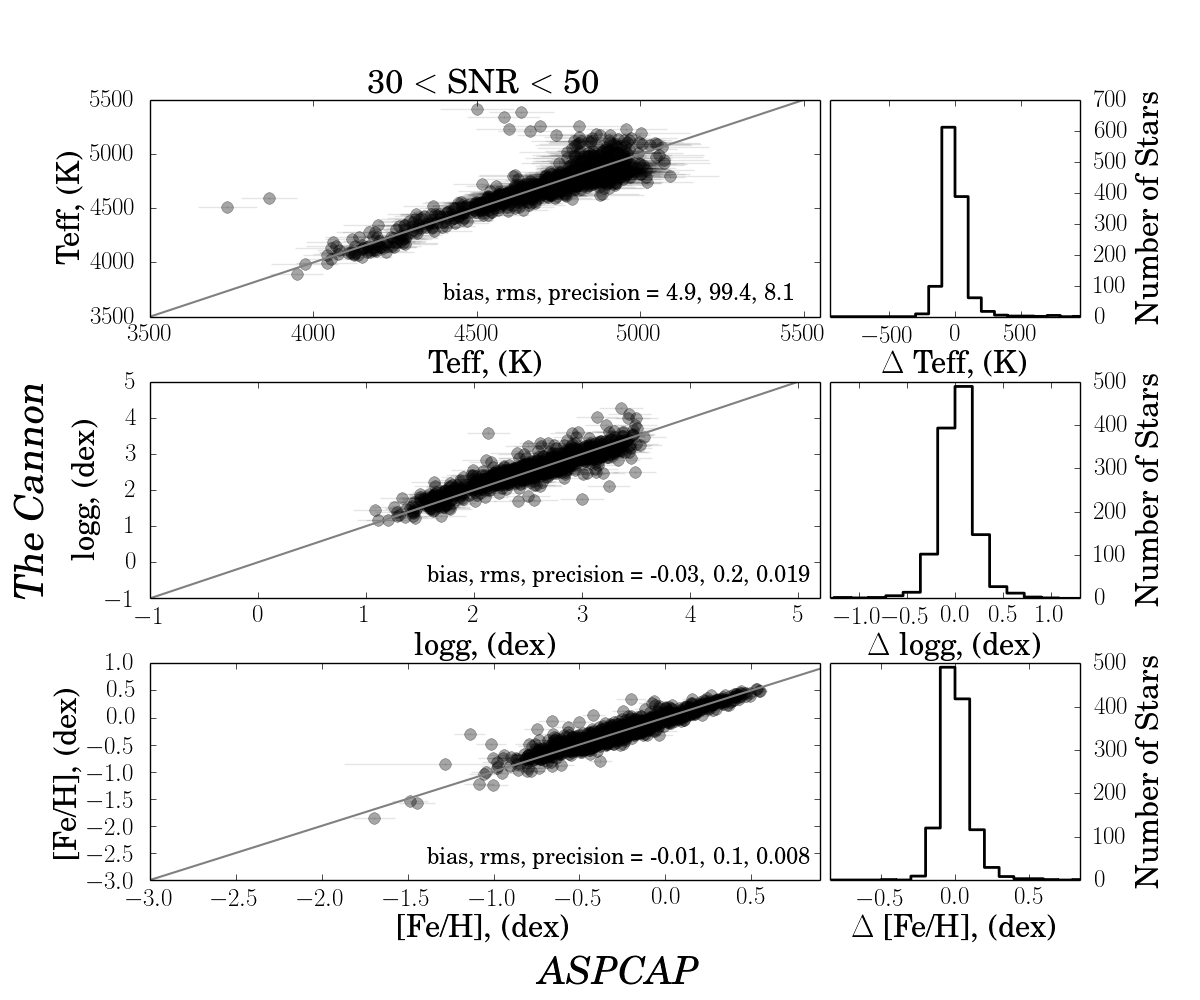}
\includegraphics[scale=0.25]{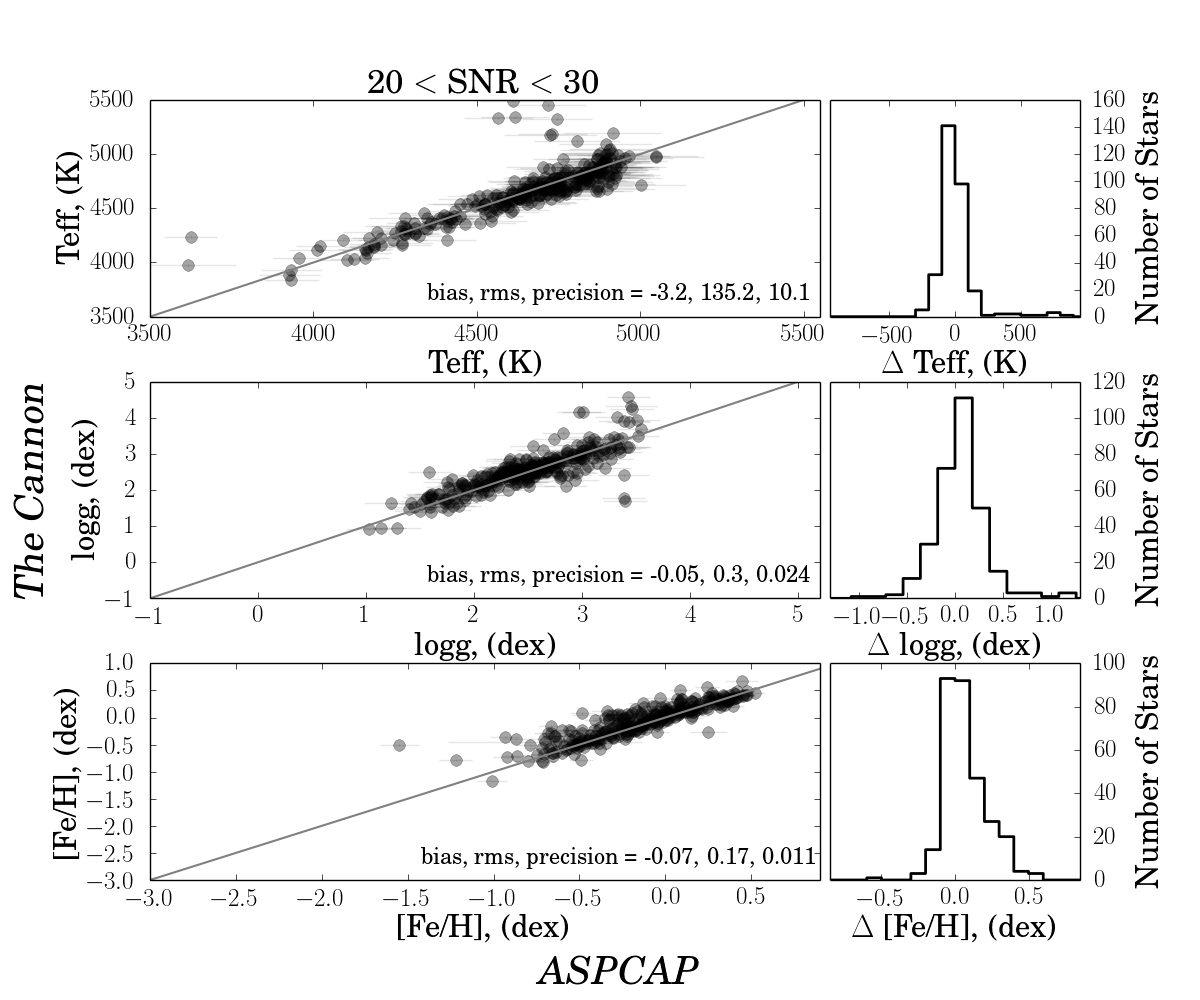}
    \caption{Illustration of \tc's ability to estimate labels for spectra of modest SNR. Shown is the comparison of \tc\ labels derived for some single visit spectra, compared to the \aspcap\ label values derived from the co-added high SNR spectra. The single vista spectra are grouped in four different regimes of SNR. There are 60 stars in the 20 $<$  SNR $<$ 30 bin, 1200 stars in the 30 $<$ SNR $<$ 50 bin, 1100 stars in the 50 $<$ SNR $<$ 100 bin and 670 stars in the 100 $<$  SNR $<$ 200 bin. Note that the rms difference between those two label estimates increases more slowly than expected from the SNR of the single visit spectra: label transfer with \tc\  therefore enables label estimates at modest SNR. Each SNR regime shows the corresponding histograms of \tc\ - \aspcap\ for each label, at right.}
\label{fig:SNR}
\end{figure}

By identifying `true' continuum pixels we have been able to implement a simple continuum-normalization that is robust across low and high SNR and that is valid across the parameter range of our training set. To examine how \tc\ performs at lower SNR, we have taken individual visits from the \apstar\ fits files, when there are $\ge$ 4 visits, and run \tc\ on a single visit spectra, when consistently continuum-normalized (\sectionname~\ref{sec:ApogeeContinuum}). Note, that we have not simply added noise to the combined DR10 spectra for our low SNR tests, which would bypass the question of how consistently the continuum can be defined at different SNR levels. Instead, we have treated single-visit spectra as (formally) independent survey objects. \figurename~\ref{fig:lowsnr} shows a comparison of a sample star for a single visit and combined visits ($>$ 4 total visits). \figurename~\ref{fig:SNR} presents the results of \tc\ compared to \apogee\ for these stars, showing \textit{only} the \apogee\ stars with errors of $<$ 150 K in \teff\ and $<$ 0.25 dex in \logg, across four SNR intervals, from 20 $<$ SNR $<$ 30 to 100 $<$ SNR $<$ 200.

These \figurenames\ illustrate that our approach to continuum normalization works well for both of these SNR regimes and is SNR independent, which is not true for a weighted-quantile normalisation. At the highest SNR (and \apogee\ estimates a upper noise floor of 200 although stars do measure above this), the rms difference between \tc\ and \aspcap\ is comparable to the \aspcap\ measurement errors, at 73K in \teff\, 0.18 dex in \logg\ and 0.11 dex in \feh. At a SNR of 30-50, the rms error increases to 100 K, 0.2 dex and 0.10 dex in \teff, \logg\ and \feh\, respectively. At an SNR of 20-30 the rms error is significantly higher and here the internal errors of \tc\ become comparable to typical minimisation methods and at SNR $<$ 20 exceed them. With this method we can return stellar parameters of \teff, \logg, \feh\ to as good a precisions as minimisation techniques ( \teff\ $<$ 100K, \logg\ $<$ 0.2 dex, \feh\ $< $0.1 dex) with an SNR of $\ge$ 25. 
 
\clearpage
\section{Discussion}

We have demonstrated with \tc\ that it is possible to label stellar
spectra from extensive homogeneous surveys with stellar parameters 
and abundances (collectively ``stellar labels''), using not physical stellar models but rather a
\emph{training set} of reference objects. These reference objects must have trustworthy
labels and spectra with the same resolution, line-spread function, and
wavelength coverage (though not necessarily SNR) as the data on the 
survey objects that require labeling.
Except for the fact that the reference objects must have been assigned
labels themselves somehow, presumably on the basis of physical models for stellar 
structure and photospheres, we do not rely on explicit stellar photosphere models
for the spectra. \tc\  is based on the premise that (continuum-normalized) spectra of stars with the same labels
look the same, and that spectra vary smoothly with changing labels. 
This makes it possible to propose a simple mathematical model for the spectrum as a function of the
labels, and fix this model in the \textit{training step}, operating on the spectra of the reference objects.
In the subsequent \textit{test step} that same model can assign labels (and their uncertainties) to all
other objects in the survey.

In a first application of \tc, on the \apogee\ DR10 data, we focused on the three most important labels, \teff, \logg, \feh , and 
derived them for essentially all of the 55,000 DR10 survey stars, based on a training step that involved only 542 reference objects, i.e. 1\% of the survey.  Remarkably, \tc 's label transfer results in stellar parameters and metallicites that are as precise and accurate as those derived from
\apogee 's pipeline \aspcap. In addition, \tc\ appears to produce -- at least in this present implementation -- plausible labels for 6000 main sequence stars in \apogee, even though only $\sim 60$ main sequence stars were used in the training step (all of which are members of the Hyades cluster). 
It is also remarkable that  the \logg - \teff\  diagram of Figure.~\ref{fig:iso2}~ shows basically no stars outside
the physically plausible regime, although \tc\ knows nothing here about stellar evolution save the training step.

Our application to \apogee\  illustrates a number of strengths and practical advantages of such an approach. First, \tc\ is computationally very fast. It  trains fast and then delivers the labels for 55,000 stars of the
 \apogee\ DR10 sample in reasonable time on a single laptop: it takes $<0.1$~s on a 2.6-GHz
intel core i7 to determine three labels for each survey star, without any
attempt at code optimization. This is because \tc\  only involves linear algebra and (in the test step) the well-behaved optimization of a few parameters with an analytic model for the spectrum. 

Second, we have explicitly demonstrated that \tc\ can deliver labels, at least these three labels, 
with nearly the same precision at much lower SNR than commonly deemed necessary. 
The \textit{rms} difference between  \aspcap\ labels from spectra with SNR$\ge150$ and \tc\ labels
for the same stars from SNR$ \sim 50$ survey spectra is only 30\% larger than the \aspcap\ error bars. 
\tc\  exploits the information at all pixels and certainly the labels \teff, \logg, \feh\  effect many different parts of each spectrum. How this SNR behaviour scales to label sets of higher dimension, encompassing, for example, individual abundances, remains to be seen. Part of the reason for this good behavior at low SNR is presumably that
\tc\ contains a generative model of the intensity or flux density.
Given labels, the model provides a Gaussian \textit{pdf} for the flux density at every wavelength.
This \textit{pdf} is convolved (trivially) with the Gaussian uncertainty
assigned to each pixel measurement in the data when the comparison is
made between the observed data in the survey object spectra with the
generative model, straightforwardly accommodating
heteroscedastic uncertainties from spectrum to spectrum.

Third, \tc\ requires and provides a continuum estimate that remains  
unbiased among spectra of different SNR. The training step of \tc\ itself identifies the 
pixels that have near-unity flux in preliminarily normalized spectra, {\it and} that show little flux variation with 
label changes. Those pixels are, conceptually and practically, good approximations to pixels to which to fit a smooth continuum. Our initial application to \apogee\ spectra indicated that biased continuum fits to spectra would be the main source of poor label estimates from lower SNR spectra using \tc , and may well also be for label estimates based on physical models. 

Our initial application of \tc\ to \apogee\  DR10 data involved a number of important approximations and illustrated several important limitations. 
We discuss some of these now, along with the benefits and costs of relaxing
them. Some of these limitations are attributable to the 
particular implementation of \tc , which is just the tip of a large iceberg of potential
methods for transferring labels from a set of reference objects to a
set of unlabelled survey objects. Other limitations are inherent to the overall approach. 
The current limitations mainly revolve around the (reference) labels on the one hand, and the choice 
of the spectral model on the other hand. 

Three important issues arise around labels. First, the reference labels are so far assumed to be perfectly known, 
but in reality are both noisy and potentially biased. In turn, we presume that we
have simply no additional information about the labels of the survey objects. Yet,
we know \emph{something} about the unlabelled stars (for example, from
photometry, and stellar evolution models). Second, any choice for the dimensionality of the label space, 3D in our sample application, will be incomplete in an astrophysical sense. Clearly, stars with identical \teff, \logg, \feh\ may have different spectra, for example,
because they differ in [$\alpha$/Fe] or $v_{rot}$.
Third, no set of reference objects will cover the label
space comprehensively, especially if one considers high-dimensional label spaces (\apogee 's DR12
published 16 labels per star!). 

The general approach to the first and second issues is to expand the scope of the model, which warrants substantive discussion. The model currently only generates spectra by providing a \textit{pdf}
over spectral pixel intensities given a set of labels.
Symbolically we could write that \tc\ in its current implementation
learns or provides a conditional \textit{pdf}
$p(f_\lambda\given\starlabelvec,\set{\theta}_\lambda, s_\lambda^2)$ (see Equation \ref{eq:like}).
Given a prior on the label space
$p(\starlabelvec)$, \tc\ could straightforwardly become a generative model of both
the spectral pixel intensities \emph{and} the labels.

This would also make it possible to learn the spectral model $\set{\theta}$ from
reference objects with noisy labels: at the moment, we effectively 
assume for the reference objects that $p(\starlabelvec_n)$ is a delta-function 
at the known labels. For noisy labels of the reference objects 
we would set $p(\starlabelvec_n)$ instead to reflect the label uncertainties. 
One would then, however, have to optimize simultaneously
$\set{\theta}_\lambda$ for \textit{all} $\lambda$ pixels and the labels $\starlabelvec_n$
for \textit{all} $n$ reference objects. Missing labels among some of the reference
objects could be treated pragmatically as simply having very large uncertainties.

Thinking of \tc\ as a model for both the spectral intensities and the labels
also shows how any (much more limited) external information on the labels of the survey
objects could be incorporated. One learns from the reference and survey objects
simultaneously (effectively, lifting the separation of training and test step),
by optimizing $\set{\theta}_\lambda$ and $\starlabelvec_n$, where the index $n$ now 
encompasses both the reference and survey sample. The difference between reference and survey objects
now simply consists of how tightly constrained their $p(\starlabelvec_n)$ is. 
For survey objects, $p(\starlabelvec_n)$ will likely be broad, for example, constraining label-combinations
to physically plausible isochrones. This would combine aspects of \tc\ with the approach
taken by \cite{SB2014}.

A generative model of both the spectra \emph{and} the labels
would in principle be much more powerful than the current generative
model of spectra alone.
That said, these joint optimizations of parameters and labels would be
expensive and multi-modal, so it might not be computationally
tractable.

In this initial implementation of \tc\ we restricted ourselves to producing only the maximum-likelihood estimates for both the $\set{\theta}_\lambda$ in the training step, and the $\starlabelvec_n$
in the test step, with label errors only coming from the inverse covariance matrix at that point.
A full inference would be expensive, especially in the test step
(labeling the survey objects); the test step model is non-linear and
inference would require sampling or harsh approximations.The correct way to proceed in the full interference case would be in a fully Bayesian framework, where test and training happen at the same time and are not separated out, as is the case now.  This is far more computationally expensive than the current approach.  Indeed, all straightforward implementations of a joint inference of parameters and labels given noisily labeled training data and unlabeled test data are computationally intractable at present.  There are promising approaches that involve either variational inference or Gibbs sampling, but developing either into a practical approach is a significant research project, not just in astrophysics but also in inference. 

In the application of \tc\ to \apogee\ we dealt with systematic errors in the reference
labels by adjusting them,  given the unphysically narrow giant branch returned for DR10 data at low \logg.  We empirically found by adopting a very naive calibration that shifts the stars to the nearest position on the isochrone from the \aspcap\ value described in \sectionname~\ref{sec:ReferenceObjects}, the stars were returned in a \teff-\logg\ space, across metallicity, in line with expectations of the physical label-space of stars. This suggests that there is some problem with the input labels in either the \teff\ or the \logg\ dimension adjusted from Kepler results in DR10. 

The issues raised above on the dimensionality and on the coverage of label space by the reference objects, are linked: 
the basic implementation of \tc\ presented here considers only three
labels (\teff, \feh, and \logg), and we know that the label-space has many more dimensions.
Conceptually, it is trivial to extend \tc\ to even much larger numbers of labels per star. 
For example, a next generation could include \alphafe\ or \xfe\
labels for elements X; but also stellar rotation, or photospheric turbulence could be labels, provided suitable sets of reference values exist.
The only limitation---and it is a \emph{substantial} limitation---is
that as the label-space grows, we presume that the training set must grow to fill it.
After all, \tc\ can only be as good as its training set.
We have shown that the set of 542 reference objects (1$\%$ of the survey) does well for three labels. However in general, the training set needs may scale up as badly as exponentially with the dimensionality of the label space. Therefore, it is at this point an open issue, to 
how many label-dimensions \tc\ continues to be useful and practicable. In expanding the list of labeled stars, one option is to identify critical targets for careful labelling, using new (possibly expensive) data and (definitely expensive) human time to obtain good labels, on the same abundance and stellar-parameter scale as the labels we already have in our limited training set.  Heuristically, we want new labeled targets to be in parts of the label space not covered (or poorly covered) by the existing training set.  More quantitatively, we could use ideas from experimental design or active learning \citep{active} to make optimized choices.  Good technology here could permit us to expand the dimensionality of the label space while growing the size of the training set as minimally and as objectively as possible.

\tc\ is related in a number of ways to supervised methods within the domain of machine learning. On the one hand, \tc\ is a pure supervised classification method:  It is trained on data and labels from a population that is assumed to have perfect labels and it is applied to data for which labels are assumed to be unknown.  However, \tc\ is also very unlike standard machine-learning methods in a critical respect:  It makes no assumption that the test data and the training data are statistically similar, or drawn from the same noise distribution.  Indeed, the main reason that the method is written as a generative probabilistic model is precisely so that it can account for the changing noise model (changing noise variances) from object to object and pixel to pixel.  Standard supervised methods from the machine-learning literature (such as Random Forest \citep{RF}, Deep Learning \citep[e.g.,][]{DL2,DL3,DL}, and Kernel Support Vector Machines \citep{SVM} do not have the property that they can account for variable noise models. These traditional machine-learning methods perform very badly as the training data become different from the test data (as they do in our SNR experiments in Section~\ref{sec:lowSNR}).

In this sense, \tc\ is less like a standard machine-learning method and more like one of the new methods being developed to account for differences between the training set and the test set.  In some sense, \tc\ is really a Transfer Learning method \citep[e.g.,][]{TL} because it learns on data with one noise model (or many noise models) and then is employed on data drawn from a new set of noise models.  In the future, as \tc\ is understood and developed further, we expect there to be enhancements that benefit from new developments in machine learning.  For example, the fact that the test set might (or does) span a different part of label space than the training data might be accommodated by ideas from Concept Drift \citep[e.g.,][]{CD} or Model Adaptation \citep[e.g.,][]{MA}, both of which are being developed precisely to account for the problem that in many real-world applications of machine learning.

An important aspect of \tc\ as presented here, concerns the spectral model itself. With the pixel-by-pixel polynomial \textit{Ansatz} for the spectral model $\set{\theta}_\lambda$
we engender two important consequences. First, we need to pick a functional form for the spectral model
(Equation \ref{eq:specmodel}), which we took empirically to be a quadratic-in-labels form of Equation \ref{eq:quadinlabels}. We arrived at that choice by empirical experimentation with this particular data set, but this choice can be generalized.
Indeed, the polynomial family is probably not the best family of
functions to be exploring, since they extrapolate badly (edge effects)
and require explicit, qualitative choices about order and cross-terms.
It is probably better to eventually move to a non-parametric form for the functions,
such as Gaussian Processes.
In this case, model complexity would be controlled by continuous
parameters and the functional form could become very complex at the pixels where the data in the
training step warrant it.
This would be a natural extension of what has been implemented here.

Second, our current \textit{Ansatz} for the spectral model treats all
pixels independently, which they plausibly are only in their noise properties.
This approximation was made to make the system fast; training
(learning) can take place at each wavelength independently and (in
principle) in parallel.
However, it is not a good approximation for many reasons.
One of these is that the finite resolution of the spectrograph
correlates nearby pixels; the generative spectral model cannot vary
substantially over wavelength differences that are far smaller than
the spectrograph resolution.
This point of prior information is not used at all in the model.

A much more complex imperfection of the independent-pixel assumption is 
 that there are multiple lines from the same element and same ionization state.
These are expected to be co-variant in any sensible model.
We do not make any use of such information; indeed no line list enters
\tc\ at any stage.
These decisions were made for good, pragmatic computational reasons.
A better model would permit itself to know about the spectrograph
resolution and either know about or discover sets of lines that vary
together.
However, any such generalization will come at substantial computational cost.

Both the application to \apogee\ data and the possibilities to apply \tc\ in a broader context, bring the question of suitable sets of reference objects into focus. Indeed, in the long run the biggest practical problem in applying \tc\ may not linked to the mathematics of the method \textit{per se}, but to the actual availability of sufficiently many and sufficiently diverse reference objects in the survey to cover label space in the training step. The most glaring issue in the \apogee\ DR10 case, even with only three labels per star, is that fact that all main sequence stars in the reference set of objects come from only one cluster, without any range in metallicity. With for example, the DR12 of \apogee\ training sets of much higher dimensionality are becoming available (especially [X/H] of individual elements). While this prospect is exciting, it will exacerbate both the question of how to make labels space coverage sufficient for the training step, and how to assert the accuracy of the training labels in the first place. 

Fortunately, there are in principle quite a number of options for picking reference objects. One could pick a subset of survey objects (sensibly covering label space) where the spectra have exceptionally high SNR, 
lending particular credence to the labels derived from physics-based models. On these, one would train the spectral model and then transfer labels to the remaining survey objects, effectively deriving most of the survey labels from the observations of highest SNR.
Alternatively, as we did here, one can choose reference objects where special circumstance (cluster membership, astroseismological information) lend particular credence to their labels. 

As labels are a property of the star, not of the data set at hand, they can come from completely independent sources of information. Even if we understood absolutely nothing about near-IR spectra, but had labels for the 542 reference objects from optical spectroscopy, we could have derived the labels for the \apogee\ DR10 stars as well as \aspcap. This leads to perhaps the most exciting long-term prospect of \tc : bringing
qualitatively different stellar surveys---surveys that use different
instruments, working in different wavelength regions at different
resolutions and SNRs---onto a consistent stellar parameter and
chemical abundance scale.
So long as different surveys can agree on benchmark stars and best
values for the stellar labels, and so long as those training sets are
large enough and span enough of the label space, \tc\ (or a future
upgrade that implements some of the ideas in this \sectionname) can be
used to ensure that all of the surveys are delivering stellar
parameters on the same system.
\tc\ will not make the data coming from any survey more
\emph{accurate}, but it might serve to make the whole industry of
stellar parameter estimation and element abundance tagging more \emph{precise} and consistent.

This prospect of survey self-labeling (for example, from high SNR to low SNR) and the prospect of cross-survey calibration brings even more urgency to assuring that sufficient calibration observations are in place and that the different major spectroscopic surveys have sufficient sample overlap.

\acknowledgements

We would like to thank Daniel Foreman-Mackey (NYU), 
Morgan Fouesneau (MPIA), Jon Holtzman (NMSU),  Keivan Stassun (Vanderbilt University), Jennifer Johnson (OSU) and David Sontag (NYU)
for valuable discussions.
DWH was partially supported by
the NSF (grant IIS-1124794), NASA (grant NNX08AJ48G), and the
Moore--Sloan Data Science Environment at NYU.
The research has received funding from the European Research Council under the European
Union's Seventh Framework Programme (FP 7) ERC Grant Agreement n.
[321035].

Funding for SDSS-III has been provided by the Alfred P. Sloan Foundation, the Participating Institutions, 
the National Science Foundation, and the U.S. Department of Energy Office of Science. The SDSS-III web site is \url{http://www.sdss3.org/}.

SDSS-III is managed by the Astrophysical Research Consortium for the Participating Institutions of the SDSS-III Collaboration
 including the University of Arizona, the Brazilian Participation Group, Brookhaven National Laboratory, Carnegie Mellon University, 
 University of Florida, the French Participation Group, the German Participation Group, Harvard University, the Instituto de Astrofisica 
 de Canarias, the Michigan State/Notre Dame/JINA Participation Group, Johns Hopkins University, Lawrence Berkeley National Laboratory, 
 Max Planck Institute for Astrophysics, Max Planck Institute for Extraterrestrial Physics, New Mexico State University, New York University, 
 Ohio State University, Pennsylvania State University, University of Portsmouth, Princeton University, the Spanish Participation Group, 
 University of Tokyo, University of Utah, Vanderbilt University, University of Virginia, University of Washington, and Yale University


\end{document}